\documentclass[12pt]{article}
\usepackage{xcolor}
\usepackage{chet}
\usepackage{textcomp}
\usepackage{cite}
\usepackage{multirow}
\usepackage{enumerate}

\usepackage{graphicx}
\usepackage{tikz}
\usepackage{subfig}

\usepackage{amsfonts,amsmath}
\usepackage{braids}
\usepackage{mathtools}
\newcommand{\bZ}{\mathbb{Z}}
\newcommand{\bF}{\mathbb{F}}
\newcommand{\bC}{\mathbb{C}}
\newcommand{\bR}{\mathbb{R}}
\newcommand{\bP}{\mathbb{P}}
\newcommand{\cO}{\mathcal{O}}
\newcommand{\cN}{\mathcal{N}}
\newcommand{\cM}{\mathcal{M}}
\newcommand{\cs}[1]{\cM_{cs}(#1)}
\newcommand{\vs}{\vspace{.3cm}}

\def\ge{{\mathfrak{e}}}
\def\gso{{\mathfrak{so}}}
\def\gsu{{\mathfrak{su}}}
\def\gsp{{\mathfrak{sp}}}
\def\gf{{\mathfrak{f}}}
\def\gg{{\mathfrak{g}}}

\preprint{NSF-KITP-16-023}

\vspace{-.2cm}
\title{\Large\bf Strong Coupling in F-theory\\ \vspace{.1cm} and Geometrically Non-Higgsable Seven-branes}

\author{James Halverson}

\affiliation{Department of Physics, Northeastern University\\
Boston, MA 02115-5000 USA \\ \vspace{.5cm} Kavli Institute for Theoretical Physics, University of California\\
Santa Barbara, CA 93106-4030 USA}

\abstract{Geometrically non-Higgsable seven-branes carry gauge sectors
  that cannot be broken by complex structure deformation, and there is
  growing evidence that such configurations are typical in F-theory.
  We study strongly coupled physics associated with these
  branes. Axiodilaton profiles are computed using Ramanujan's theories
  of elliptic functions to alternative bases, showing explicitly that
  the string coupling is $O(1)$ in the vicinity of the brane; that it
  sources nilpotent $SL(2,\bZ)$ monodromy and therefore the associated
  brane charges are modular; and that essentially all F-theory
  compactifications have regions with order one string coupling.  It
  is shown that non-perturbative $SU(3)$ and $SU(2)$ seven-branes are
  related to weakly coupled counterparts with D7-branes via
  deformation-induced Hanany-Witten moves on $(p,q)$ string junctions
  that turn them into fundamental open strings; only the former may
  exist for generic complex structure.  D3-brane near these and the
  Kodaira type II seven-branes probe Argyres-Douglas theories. The BPS
  states of slightly deformed theories are shown to be dyonic string
  junctions.}

\begin{document}
\maketitle

\toc

\clearpage
\newsec{Introduction}[SecIntro]

\vspace{-.2cm}

Gauge sectors can arise along coincident seven-branes in type IIB and
F-theory \cite{V1} compactifications, in which case the splitting of
branes gives rise to spontaneous symmetry breaking via the Higgs
mechanism. This phenomenon is well-known in simple examples in flat space,
but it generalizes to other examples, as well.

For example, in the geometric F-theory description of such setups, seven-brane
positions and splitting are controlled by the complex structure of a Calabi-Yau
elliptic fibration $X\xrightarrow{\pi}B$, where $B$ are the extra spatial
dimensions, and in some cases there are
complex structure deformations that break the gauge group $G$ to a
subgroup. If the deformation is small the branes only split a small amount and the massive W-bosons of the
broken theory are $(p,q)$ string junctions
\cite{GaberdielZwiebach,DeWolfe:1998zf}; connections between
deformations, junctions, and Higgsing have been explored in recent
physics works \cite{Grassi:2013kha,Grassi:2014sda} and rigorous
mathematical proofs \cite{Grassi:2014ffa}. 

In certain cases there exists no non-abelian gauge symmetry for generic
complex structure, i.e. the branes are generically split. It is natural to wonder, then,
whether moduli stabilization fixes vacua on subloci in moduli space with
gauge symmetry, and whether cosmology prefers such vacua. In fact,
recent estimates of flux vacua
\cite{Braun:2014xka,Watari:2015ysa} show that obtaining gauge symmetry
on seven-branes by specialization in moduli space is statistically
very costly. Specifically, the number of flux vacua that exist on 
subloci on moduli space that exhibit non-abelian gauge symmetry is exponentially
suppressed relative to  those on generic points in complex structure moduli space.
For spaces $B$ that have no gauge symmetry for generic complex structure,
obtaining gauge symmetry has a high price.

However, it has been known for many years \cite{MV} that for some spaces $B$ 
there are no
complex structure deformations that break $G$, in which case the
theory exhibits seven-branes with gauge group $G$ for
generic complex structure. This could be important for moduli
stabilization  and for addressing the prevalence of symmetry in the
landscape
\cite{GrassiHalversonShanesonTaylor2014}.  These have been called
non-Higgsable seven-branes and sometimes many such intersecting branes
exist, giving non-Higgsable clusters. This name is particularly apt in
six-dimensional compactifications, where the only known source of symmetry
breaking is complex structure deformation, so the low energy theory
cannot be Higgsed. There are other sources of symmetry breaking in four dimensional
compactifications, such as flux and T-branes \cite{Cecotti:2010bp}, so that the non-Higgsable
seven-branes are more appropriately called \emph{geometrically} non-Higgsable. Having
stated the caveats, we will henceforth use non-Higgsable, for brevity.

Based on a number of works
\cite{Morrison:2012js,Morrison:2012np,Anderson:2014gla,GrassiHalversonShanesonTaylor2014,Morrison:2014lca,Halverson:2015jua,Taylor:2015ppa}
in recent years, there is growing evidence
\cite{Morrison:2012js,GrassiHalversonShanesonTaylor2014,Halverson:2015jua,Taylor:2015ppa}
that non-Higgsable seven-branes and non-Higgsable clusters are generic
in six- and four-dimensional compactifications of F-theory.  This
evidence arises from both abstract argumentation and large datasets, as
will be reviewed in section \ref{sec:review}. 

To first approximation,
seven-brane properties in F-theory are determined by the structure of
the so-called Kodaira singular fiber over the seven-brane in the elliptic
fibration, and the non-Higgsable seven-branes always have
Kodaira fibers of type $II, III, IV, I_0^*, IV^*$, $III^*,$ or $II^*$.
Therefore, properties that are true of the seven-branes associated to
these fibers are also true of non-Higgsable clusters. Though
we will derive general results for any seven-branes with these fibers,
the results will also hold for non-Higgsable seven-branes.

Motivated by the genericity with which non-Higgsable seven-branes appear, the purpose of this paper is to study them
from a number of points of view. We will focus on
the strongly coupled physics that exists in the vicinity of the brane.

First, in section \ref{sec:axiodilaton}, we will study
the axiodilaton $\tau=C_0+ie^{-\phi}$ explicitly
as a function of coordinates on $B$. In particular, the
variation of the string coupling $g_s$ over the extra
dimensions of space will be determined. To do this,
we will use Ramanujan's theories of elliptic
functions to alternative bases, which will 
allow for the explicit inversion of $J$-function
of the fibration to obtain $\tau$. We will study a number
of concrete examples, and will also show that there is
essentially always a region in $B$ with $O(1)$ $g_s$
in compactifications with seven-branes.

Next, in section \ref{sec:su3su2} we will study non-perturbative
seven-branes that realize $SU(3)$ and $SU(2)$;
these are the only geometric $SU(N)$ groups that may exist for generic complex structure. The massless W-bosons of these
theories are shrunken $(p,q)$ string junctions. We relate
the non-perturbatve realizations to the perturbative D7-brane
description by explicit deformations, and find that 
in such a limit the
$(p,q)$ string junctions undergo a Hanany-Witten move
that turns them back into fundamental strings.

Another interesting phenomenon is that theories with
non-Higgsable seven-branes are sometimes required to have
three seven-branes intersecting in codimension two in $B$,
rather than the expected two. We study this in generality
in \ref{sec:extra branes} and study associated matter
representations at these unusual enhancement points.

Finally, in section \ref{sec:D3 probes} we will study
D3-brane probes of certain non-Higgsable seven-branes. 
D3-branes near these realize Argyres-Douglas theories,
and using BPS conditions of \cite{DeWolfe:1998bi} we will
demonstrate that the BPS states of the D3-brane theory
near slightly deformed seven-branes are string junctions.

\section{Review of Geometrically Non-Higgsable Seven-branes}
\label{sec:review}

We will study seven-branes using their geometric description in
F-theory.  There the axiodilaton $\tau = C_0 + i\, e^{-\phi}$ of the type IIB
theory is considered to be the complex structure modulus of an
auxiliary elliptic curve which is fibered over the compact extra dimensional space
$B$. Such a structure is determined by a Calabi-Yau fourfold $X$
which is elliptically fibered $\pi:X\rightarrow B$. An elliptic fibration with section
is birationally equivalent \cite{Na88} to a Weierstrass model
\eqn{
y^2 = x^3 + f\, x + g
}[]
where $f$ and $g$ are sections of $K_B^{-4}$ and $K_B^{-6}$, respectively, with
$K_B$ the anticanonical bundle on $B$. The fibers $\pi^{-1}(p)$ are smooth
elliptic curves for any point $p$ which is not in the discriminant
locus
\eqn{
\Delta = 4\,f^3 + 27\, g^2 = 0.
}[]
On the other hand if $p$ is a generic point in the codimension one locus 
$\Delta = 0$, then $\pi^{-1}(p)$ is one of the singular fibers classified by
Kodaira \cite{MR0187255,MR0205280,MR0228019}.

Seven-branes are located along $\Delta = 0$. Their precise nature
depends on the structure of $f$ and $g$ and therefore also $\Delta$, which may be an irreducible
effective divisor or comprised of components
\begin{equation}
\Delta = \prod_i \Delta_i.
\end{equation}
Taking a loop around $\Delta$ or any component $\Delta_i=0$ induces an
$SL(2,\bZ)$ monodromy on the associated type IIB supergravity
theory. The action on $\tau$ is
\begin{equation}
\tau \mapsto \frac{a\tau + b}{c\tau + d}, \qquad \qquad M = \begin{pmatrix}a & b \\ c & d\end{pmatrix}\in SL(2,\bZ).
\end{equation}
Seven-brane structure is determined by the Weierstrass model according
to the order of vanishing of $f$, $g$, and $\Delta$ along the
seven-brane. Often in this paper some $\Delta_i=z^N$, and therefore we
will denote the associated orders of vanishing as $ord_z(f,g,\Delta)$
as a three-tuple or in terms of the individual orders $ord_z(f)$,
$ord_z(g)$, and $ord_z(\Delta)$. From this data the singularity type,
$SL(2,\bZ)$ monodromy, and non-abelian symmetry algebra (up to outer
monodromy) can be determined; see Table \ref{tab:fibs}.  This is the
geometric symmetry group, henceforth symmetry group or gauge group, along the
seven-brane in the absence of symmetry-breaking $G$-flux.  The
structure of $\Delta$ is determined by $f, g$ and there is a moduli space of such choices that corresponds to
the complex structure of $X$. Gauge sectors along seven-branes can be
engineered by tuning $f$ and $g$ relative to their generic structures.

\begin{table}[thb]
  \centering
\scalebox{.8}{
 \begin{tabular}{|c c c c c c c c|}
\hline
Type &
$ord_z(f)$ &
$ord_z(g)$ &
$ord_z(\Delta)$ &
singularity & nonabelian symmetry algebra & monodromy & order \\ \hline \hline 
$I_0$&$\geq $ 0 & $\geq $ 0 & 0 & none & none  & $\begin{pmatrix} 1 & 0 \\ 0 & 1\end{pmatrix}$ & $1$\\ 
$I_n$ &0 & 0 & $n \geq 2$ & $A_{n-1}$ & $\gsu(n)$  or $\gsp(\lfloor
n/2\rfloor)$& $\begin{pmatrix} 1 & n \\ 0 & 1\end{pmatrix}$ & $\infty$\\
$II$ & $\geq 1$ & 1 & 2 & none & none & $\begin{pmatrix} 1 & 1 \\ -1 & 0\end{pmatrix}$ & $6$\\
$III$ &1 & $\geq 2$ &3 & $A_1$ & $\gsu(2)$ & $\begin{pmatrix} 0 & 1 \\ -1 & 0\end{pmatrix}$ & $4$\\
$IV$ & $\geq 2$ & 2 & 4 & $A_2$ & $\gsu(3)$  or $\gsu(2)$& $\begin{pmatrix} 0 & 1 \\ -1 & -1\end{pmatrix}$ & $3$\\
$I_0^*$&
$\geq 2$ & $\geq 3$ & $6$ &$D_{4}$ & $\gso(8)$ or $\gso(7)$ or $\gg_2$ & $\begin{pmatrix} -1 & 0 \\ 0 & -1\end{pmatrix}$ &$2$\\
$I_n^*$&
2 & 3 & $n \geq 7$ & $D_{n -2}$ & $\gso(2n-4)$  or $\gso(2n -5)$ & $\begin{pmatrix} -1 & -n \\ 0 & -1\end{pmatrix}$ & $\infty$\\
$IV^*$& $\geq 3$ & 4 & 8 & $E_6$ & $\ge_6$  or $\gf_4$& $\begin{pmatrix} -1 & -1 \\ 1 & 0\end{pmatrix}$ & $3$\\
$III^*$&3 & $\geq 5$ & 9 & $E_7$ & $\ge_7$ & $\begin{pmatrix} 0 & -1 \\ 1 & 0\end{pmatrix}$ & $4$\\
$II^*$& $\geq 4$ & 5 & 10 & $E_8$ & $\ge_8$ & $\begin{pmatrix} 0 & -1 \\ 1 & 1\end{pmatrix}$ & $6$\\
\hline
non-min &$\geq 4$ & $\geq6$ & $\geq12$ & \multicolumn{4}{c|}{does not
appear for supersymmetric vacua} \\
\hline
 \end{tabular}}
  \caption{The Kodaira fibers, along with their orders of vanishing in a Weierstrass model, singularity type, possible nonabelian
symmetry algebras, $SL(2,\bZ)$ monodromy, and monodromy order.}
  \label{tab:fibs}
\end{table}

\vspace{1cm} Let us now turn to geometrically non-Higgsable
seven-branes. Physically, this means that there are no directions in the
supersymmetric moduli space that break the gauge group on the
seven-branes by splitting them up.  Mathematically, a geometrically
non-Higgsable seven-brane along $z=0$ exists when
\begin{equation}
\Delta = z^N \, \tilde \Delta
\end{equation}
for any choice of $f$ and $g$, i.e. for a generic point in the complex
structure moduli space of $X$, henceforth $\cs{X}$. For $N>2$ the seven-brane
carries a non-trivial gauge group $G$. It is often possible that by tuning $f,g$ to a
subvariety $L\subset \cs{X}$ the discriminant $\Delta$ is proportional
to $z^{M>N}$ and the gauge group along the seven-brane at $z=0$ is
enhanced to $G'\supset G$. There may be many such loci $L_i$ in $\cs{X}$. The statement that a
non-Higgsable seven-brane exists for generic complex structure moduli is the statement
that it exists for any complex structure in $\cs{X}\setminus
\{\bigcup_i L_i\}$, which is the bulk of $\cs{X}$ since each $L_i$
has non-trivial codimension. Often the discriminant is of the form
\begin{equation}
\Delta = \tilde \Delta \,\,\prod_i z_i^{N_i}
\end{equation}
for generic complex structure, in which case there is a non-Higgsable
seven-brane along each locus $z_i=0$. They may intersect, giving rise
to product group gauge sectors with jointly charged matter that arise
from clusters of intersecting seven-branes. These are referred to as
non-Higgsable clusters \cite{Morrison:2012np,Morrison:2012js}. 

The
possible gauge groups that may arise along a non-Higgsable
seven-brane are $E_8, E_7, E_6, F_4, SO(8), SO(7), G_2, SU(3),$ and
$SU(2)$ and there are five two-factor products with jointly charged matter
that may arise. In particular, note that $SU(5)$ and $SO(10)$, which arise
from $I_5$ and $I_1^*$ fibers, may never be non-Higgsable; more generally,
this is true of seven-branes with fibers $I_n$ and $I_{n>0}^*$. This is easy
to see in the $I_n$ case. Such a model has $ord_z(f,g,\Delta)=(0,0,n)$,
and $f\mapsto (1+\epsilon) f$ for $\epsilon \in \bC^*$ is a symmetry breaking complex
structure deformation that always exists, by virtue of the model existing in the first place.
Similar arguments exist for $I_{n>0}^*$ fibers.

The name ``non-Higgsable clusters'' is a suitable name in
six-dimensional compactifications of F-theory, since there the
associated six-dimensional gauge sectors do not have any symmetry
breaking flat directions in the supersymmetric moduli space, as
determined by $\cs{X}$ and also the low energy degrees of
freedom. However, in four dimensional compactifications there are
other effects such as T-branes \cite{Cecotti:2010bp} that may break
the gauge group, so that \emph{geometrically non-Higgsable} is a more
accurate name.  Furthermore, if $\Delta \sim z^2$ for a generic $p\in
\cs{X}$ then $G=\emptyset$ even though there is a divisor $z=0$ in $B$
that is singular, and sometimes a codimension two locus $C$ may be
singular for generic moduli even if it is not contained in a
non-Higgsable seven-brane. Both have been referred to as
``non-Higgsable structure'' \cite{Halverson:2015jua} even though there
is no associated gauge group. The general feature is the existence of
singular structure for generic complex structure moduli, and aside
from these two caveats there is a gauge group on a seven-brane that
cannot be spontaneously broken by a complex structure deformation.

Though not named as such at the time, the first F-theory
compactifications with non-Higgsable seven-branes appeared in
\cite{Morrison:1996pp}. These examples have six non-compact dimensions
and four compact dimensions $B_2$ with $B_2=\bF_n$, and there is a
non-Higgsable seven-brane on the $-n$ curve in $\bF_n$ for $n>2$. The
complete set of non-Higgsable clusters and seven-branes that may arise
in six-dimensional compactifications were classified in
\cite{Morrison:2012np} and the examples with toric $B_2$ were
classified in \cite{Morrison:2012js}.  In the latter, all but $16$ of
the $61,539$ examples exhibit non-Higgsable clusters or seven-branes,
and the $16$ that do not are weak Fano varieties, i.e. varieties
satisfying $-K\cdot C\geq 0$ for any holomorphic curve $C$. In all cases in
six dimensions the reason for geometric non-Higgsability is immediately evident
in the low energy gauge theory: either there is no matter or there is not enough
matter to allow for Higgsing consistent with supersymmetry, due to having a half hypermultiplet
in a pseudoreal representation.

In examples with four non-compact dimensions the extra dimensions of
space are a complex threefold $B_3$ and there are additional
non-Higgsable clusters and structures that do not appear in six
dimensions, including for example loops \cite{Morrison:2014lca} and
the gauge group $SU(3)\times SU(2)$
\cite{GrassiHalversonShanesonTaylor2014}. In the latter case the
matter content matches the non-abelian structure of the standard
model. A classification \cite{Halverson:2015jua} of $B_3$ that are
$\bP^1$-bundles over certain toric surfaces has non-Higgsable clusters for
$98.3\%$ of the roughly $100,000$ examples with over $500$ examples with an $SU(3)\times SU(2)$
sector. A broader exploration of toric $B_3$ using Monte Carlo techniques \cite{Taylor:2015ppa}
has non-Higgsable structure for all $B_3$ after an appropriate ``thermalization,'' and approximately $76\%$
of the examples have a non-Higgsable $SU(3)\times SU(2)$ sector. Non-Higgsable clusters also appear in
the F-theory geometry with the largest number of currently known flux vacua \cite{Taylor:2015xtz},
where vacuum counts were estimated using techniques of Ashok, Denef, and Douglas \cite{Ashok:2003gk,Denef2004}. 
It is not clear whether cosmological evolution prefers the special vacua associated with a typical $B_3$, perhaps
characterized by \cite{Taylor2015}, or the typical vacua associated with a special base $B$ that gives the largest number
of flux vacua, which may be $B_{max}$ of \cite{Taylor:2015xtz}. Needless to say, this is a fascinating question moving forward.

What is becoming clear is that non-Higgsable clusters and structure
play a very important role in the landscape of F-theory
compactifications. It has become common to say that non-Higgsable
clusters are doubly generic. The first is a strong sense: for fixed
$B$, having a non-Higgsable cluster means that there is a non-trivial seven-brane
for generic points in $\cs{X}$. The second is in a weaker, but still compelling, sense:
there is growing evidence that generic extra dimensional spaces $B$ give rise to
non-Higgsable clusters or structure. One line of evidence is in the large datasets cited above.
Another is the argument of \cite{Halverson:2015jua}: if there is a curve $C \subset B$ with $-K\cdot C<0$
then $-K$ contains $C$ and $C$ sits inside the discriminant locus, giving non-Higgsable structure on $C$.
Such $B$ are ones that are not weak Fano, and it is expected that a generic algebraic surface or threefold
is of this type. In particular, there are only $105$ topologically distinct Fano threefolds.

\vspace{1cm}
In this work we will study the strongly coupled physics associated to fibers that
can give rise to geometrically non-Higgsable seven-branes. As such, the analyses of this
paper include, but are not limited to, F-theory compactifications with non-Higgsable
seven-branes. These fibers are
\begin{equation}
II, III, IV, I_0^*, IV^*, III^*, II^*,
\end{equation}
and any seven-brane with one of these has an associated $SL(2,\bZ)$ monodromy
matrix $M$ that is nilpotent, i.e. $M^k=1$ for some $k$. $M_{I_0^*}=-1$ which acts trivially on
$\tau$, indicating that this configuration is uncharged, in agreement with the fact that it is
$4$ $D7$-branes on top of an $O7$ plane from the type IIB point of view. The rest act non-trivially on
$\tau$ but the theory comes back to itself after taking $k$ loops around the seven-brane; the seven-brane
charges are nilpotent. Though our analyses apply more broadly,
they are of particular interest given the prevalence of non-Higgsable clusters in the landscape.

\section{Axiodilaton Profiles and Strong Coupling}
\label{sec:axiodilaton}

The primary difference between F-theory and the weakly coupled type
IIB string is that the axiodilaton $\tau = C_0 + i e^{-\phi}$ varies
over $B$ in F-theory, and therefore so does the string coupling $g_s =
e^{\langle \phi \rangle}$. The behavior of $\tau$ near seven-branes affects gauge theories on seven-branes,
as well as three-brane gauge theories or string scattering in the vicinity of seven-branes.
In his seminal works \cite{MR0187255,MR0205280,MR0228019} Kodaira computed $\tau$ locally near
seven-branes in elliptic surfaces.

In this section we will study axiodilaton profiles via their relation to the Klein $j$-invariant
of an elliptic curve for elliptic fibrations of arbitrary dimension.
We will normalize the $j$-invariant in a standard way by $J:= j /1728$, and in the
case of a Weierstrass model we have
\begin{equation}
J = \frac{4f^3}{\Delta} \qquad \text{where} \qquad \Delta = 4f^3 + 27g^2.
\end{equation}
In this formulation the $J$-invariant depends on base coordinates
according to the sections $f$ and $g$ of the Weierstrass
model. However, $J$ also depends on the ratio of periods of the
elliptic curve $\tau = \frac{\omega_1}{\omega_2}$ where $\tau$ is the
value of the axiodilaton field at each point in $B$. Thus, if $z$ is a
base coordinate we compute $J=J(z)$ directly from the Weierstrass
model, but this can also be thought of as $J = J(\tau(z))$. By
inverting the $J$-function, we will determine the axiodilaton profile
and study it in the vicinity of geometrically non-Higgsable seven-branes.
We will also demonstrate that F-theory compactifications generically exhibit
regions with $O(1)$ string coupling and recover classic results from the perturbative type IIB theory.

\begin{table}[thb]
  \centering
  \begin{tabular}{|ccc|} \hline
    Fiber & $J$ & $J|_{z=0}$ \\ \hline \hline
 $II$ & $\frac{z}{A+z}$& $0$ \\ \hline
 $III$ & $\frac{1}{1+Az}$& $1$ \\ \hline
 $IV$ & $\frac{z^2}{A+z^2}$& $0$ \\ \hline
 $I_0^*$ & $\frac{1}{1+A}$& $\frac{1}{1+A}$ \\ \hline
 $IV^*$ & $\frac{z}{A+z}$& $0$ \\ \hline
 $III^*$ & $\frac{1}{1+Az}$& $1$ \\ \hline
 $II^*$ & $\frac{z^2}{A+z^2}$& $0$ \\ \hline
  \end{tabular}
  \caption{The $J$-invariant for seven-branes associated to geometrically non-Higgsable clusters, expressed in a way that is particularly useful for a local analysis near the seven-brane. Here $f=z^n\, F$ and  $g=z^m\, G$ with $A= 27G^2/4F^3$.}
  \label{tab:Jw}
\end{table}

Though there are seven Kodaira fiber types that may give rise to geometrically
non-Higgsable seven-branes, $\{II,III,IV,I_0^*,IV^*,III^*,II^*\}$, some
have the same $J$-invariant and $\tau$ in the vicinity
of the brane. In each case the Weierstrass model takes the form
\begin{equation}
f = z^n \, F, \qquad g=z^m\, G, \qquad \Delta = z^{min(3n,2m)}\, \tilde \Delta,
\end{equation}
and the $J$-invariant takes a simple form. Near a generic region of the
seven-brane on $z=0$ both $F$ and $G$
are non-zero, and therefore so is $A \equiv 27G^2 / 4F^3$. The possibilities
are computed in Table \ref{tab:Jw} and the redundancies are \cite{MR0187255,MR0205280,MR0228019}
\begin{equation}
J_{II} = J_{IV^*}, \qquad J_{III} = J_{III^*}, \qquad J_{IV} = J_{II^*}.
\end{equation}
This result may seem at odds with the monodromy order for these
Kodaira fibers displayed in Table \ref{tab:fibs}, since the type $II$
and $II^*$ fibers have order $6$ whereas the type $IV$ and $IV^*$
fibers have order $3$. The resolution is that, though the monodromy
associated with type $II$ and $II^*$ fibers is $6$, $M_{II}^3 =
M_{II^*}^3=-I$, where $I$ is the identity matrix, so that the type $II$,
$II^*$, $IV$, and $IV^*$ fibers all induce an order $3$ action on $\tau$.

There are some special values for $\tau(J)$ that we will see arise in inverting
$J$,
\begin{equation}
\tau(0)=e^{\pi i/3}, \qquad \tau(1) = i,
\end{equation}
up to an $SL(2,\bZ)$ transformation. These values correspond to $g_s = \frac{2}{\sqrt{3}}$
and $g_s=1$, and it is important physically that these cannot be lowered
by an $SL(2,\bZ)$ transformation. Mapping $\tau\mapsto \tau':=\frac{a\tau + b}{c\tau + d}$
by an arbitrary $SL(2,\bZ)$ transformation for $\tau=e^{\pi i/3}$ and $\tau =i$, respectively,
we have new string coupling constants
\begin{equation}
g_s'=(c^2+cd+d^2)\frac{2}{\sqrt{3}}\geq \frac{2}{\sqrt{3}}, \qquad g_s'=(c^2+d^2)\geq 1,
\label{eq:cantgoweak}
\end{equation}
showing that the string couplings
with these two values of $\tau$ cannot be lowered by a global $SL(2,\bZ)$ transformation.

For each case in Table \ref{tab:Jw} we will invert $J$ to solve for
$\tau$.

\subsection{Inverting the $J$-function and Ramanujan's Alternative Bases}

There is a nineteenth century procedure for inverting the $J$-function that is due
to Jacobi. Recall that the $j$-invariant satisfies
\begin{equation}
j(q) = \frac{1}{q} + 744 + 196884 \,\, q + \dots 
\end{equation}
in terms of $q=e^{2\pi i  \tau}$. Jacobi's result relates $j$ to $q$ via hypergeometric
functions, which then allows for the computation of $\tau$ by taking a logarithm.
The result is
\begin{equation}
\tau = i\,\,\, \frac{_2F_1(\frac12,\frac12;1;1-x)}{_2F_1(\frac12,\frac12;1;x)}, \qquad J = \frac{4\, (1-x(1-x))^3}{27\, x^2(1-x)^2}, 
\label{eq:tauJJacobi}
\end{equation}
in terms of the hypergeometric function $_2F_1(a,b,c;x)$. For $|x|<1$ it satisfies
\begin{equation}
_2F_1(a,b;c;x) = \sum_{n=0}^\infty \frac{(a)_n(b)_n}{(c)_n \, n!} x^n,
\end{equation}
where $(a)_n=a(a+1)(a+2)\dots(a+n-1)$ for $n\in \bZ^+$ is the
Pochhammer symbol. For a particular value of $J$, then, six values of
$\tau$ are obtained by solving the sextic in $x$, and these are
related to one another by $SL(2,\bZ)$ transformations.

Much progress was made in the theory of elliptic functions at the beginning of the
$20^{\text{th}}$ century by Ramanujan, who recorded his theorems in notebooks \cite{MR0099904}
that were dense with results. In one, he claimed that there are similar inversion formulas
where the base $q$ is not
\begin{equation}
q = exp\left({-\pi \frac{_2F_1(1/2,1/2,1,1-x)}{_2F_1(1/2,1/2,1,x)}}\right)
\end{equation}
as it was for Jacobi, but is instead one of
\begin{align}
q &= exp\left({-\frac{2\pi}{\sqrt{3}} \frac{_2F_1(1/3,2/3,1,1-x)}{_2F_1(1/3,2/3,1,x)}}\right), \nonumber \vspace{2cm} \\ 
q &= exp\left({-\frac{2\pi}{\sqrt{2}} \frac{_2F_1(1/4,3/4,1,1-x)}{_2F_1(1/4,3/4,1,x)}}\right), \nonumber \\
q &= exp\left({-2\pi \frac{_2F_1(1/6,5/6,1,1-x)}{_2F_1(1/6,5/6,1,x)}}\right).
\label{eq:altbaseinit}
\end{align}
There has been significant progress \cite{MR1311903,MR1117903,MR1071759,MR1010408,MR1237931,MR1243610,MR1825995,MR3107523,Cooper2009} in the study of Ramanujan's theories of elliptic functions
to these alternative bases in recent years, including rigorous proofs of many of Ramanujan's results.
Practically, these different theories give different ways to study $\tau$.

In studying the relationship between $J$, $\tau$, and Ramanujan's alternative theories, we will
utilize the notation of Cooper \cite{Cooper2009}. The alternative bases satisfy
\begin{equation}
q_r := exp\left(\frac{-2\pi}{\sqrt{r}}\frac{_2F_1(a_r, 1-a_r;1;1-x_r)}{_2F_1(a_r,1-a_r;1;x_r)}\right)
\label{eq:altq}
\end{equation}
where $a_1=\frac16$, $a_2=\frac14$, and $a_3=\frac13$ for $r=1,2,3$ reproduce \eqref{eq:altbaseinit},
where the $J$ invariant satisfies
\begin{equation}
J = \frac{1}{4\,x_1(1-x_1)} = \frac{(1+3x_2)^3}{27\,x_2(1-x_2)^2}=\frac{(1+8x_3)^3}{64\,x_3(1-x_3)^3}.
\label{eq:JRam}
\end{equation}
For any value of $J$ one may then solve either Jacobi's sextic (\ref{eq:tauJJacobi}) or the quadratic,
cubic, or quartic in \eqref{eq:JRam}. Other inversion methods also exist, but we will not use them.

\vspace{.5cm}
We utilize these methods to study elliptic fibrations,  beginning with general statements
and then proceeding to the study of examples near geometrically non-Higgsable seven-branes.

Consider a Weierstrass model, where $J=4f^3/\Delta$. In a neighborhood of a seven-brane
on $z=0$ one can compute the $SL(2,\bZ)$ monodromy $M$ of the seven-brane by taking a small
loop around the seven-brane, which includes an action on $\tau$. Recalling that
\begin{equation}
M^k(\tau) = \tau
\end{equation} 
for some $k$ for any geometrically non-Higgsable seven-brane, one would like to verify $k$
directly by inverting the $J$-function. We will see that some of Ramanujan's theories give
rise to $k$ element sets of $\tau$ values that are permuted by the monodromy, where $k$ is the
order of $x_i$ in \eqref{eq:JRam}.

Since each solution for $x_i$ determines a value of $\tau$ directly
via $q$ in \eqref{eq:altq}, let us solve for $x_i$ in terms of $J$.
In the quadratic case we have
\begin{equation}
x_1 = \frac{1\pm \sqrt{1-1/J}}{2}
\label{eq:xinJquad}
\end{equation}
which is degenerate for $J=1$ and is not well defined for $J=0$. Aside from $I_0^*$, these
are $J$-invariants associated with geometrically non-Higgsable seven-branes. 

In the cubic case we solve the equation
\begin{equation}
27(J-1) \, x_2^3-27(J+2) \, x_2^2+ 9(3J-1) \, x_2-1 = 0
\end{equation}
to obtain $x_2$. 
Rather than solving this cubic exactly, let study the behavior of this
cubic near $J=0$ and $J=1$, since this is the relevant structure near non-Higgsable seven-branes.
Expanding around $J=0$ by taking $J=\delta J_0 \ll 1$ the three solutions for $x_2$ near $J=0$ are 
\begin{equation}
x_2 = -\frac{1}{3}-\frac{2\sqrt[3]{2}\,}{3} e^{2\pi i n/3} \,\delta J_0^{1/3}+O(\delta J_0^{2/3}), \qquad n\in\{0,1,2\}  
\label{eq:x2nearJis0}
\end{equation}
and we see that the three roots are permuted by an order three monodromy upon taking a small
loop around $J=0$. Near $J=1$ we take $J = 1+\delta J_1$ with $\delta J_1 \ll 1$ and compute
\begin{equation}
x_2 = \frac{3}{\delta J_1} + \frac{16}{9} - \frac{16\, \delta J_1}{81} + O(\delta J_1^2), \qquad
x_2 = \frac{1}{9} \pm \frac{8\sqrt{\delta J_1}}{27\sqrt{3}} + \frac{8 \delta J_1}{81} + O(\delta J_1^{3/2}) 
\label{eq:x2nearJis1}
\end{equation}
from which we see that one of the solutions goes to infinity at $J=1$ (as is expected since the cubic
reduces to a quadratic when $J=1$), whereas the other two are permuted by an order two monodromy around $J=0$.

Therefore, we will study seven-brane theories with $J=1$ $(J=0)$ with Ramanujan's theory where
$J$ is quadratic (cubic) in $x_1$ ($x_2$), solving for $\tau$.

\subsection{Warmup: Reviewing Weakly Coupled Cases}

Before proceeding to the interesting seven-brane structures that may be non-Higgsable, all of which have
finite $J$-invariant, let us consider the seven-branes that may appear in the weakly coupled type IIB
theory, which have $J=\infty$.

Let us begin with the case of $n$ coincident $D7$-branes, which in F-theory language have a Kodaira
fiber $I_n$.  The Weierstrass model takes takes the form
\begin{equation}
f = F, \qquad g = G, \qquad \Delta = z^n \tilde \Delta,
\end{equation}
where we have used our common notation of inserting $F$ in $f$ (and $G$ in $g$) even though $f,g\sim z^0$
in this case, and $z$ does not divide $\tilde \Delta$. Instead, $F$ and $G$ must be tuned to ensure the form
of $\Delta$. The $J$-invariant is
\begin{equation}
J(I_n) = \frac{4F^3}{z^n\tilde \Delta}=:\frac{C}{z^n},
\end{equation}
and we can see that, indeed, $J=\infty$ at $z=0$. Solving the theory where $J$ is a quadratic
in $x_1$ gives
\begin{equation}
x_1 = \frac12 \pm \frac12 \sqrt{1-\frac{z^n}{C}}=: \alpha_\pm,
\end{equation}
and then using equation (\ref{eq:altq}) to compute $\tau(\alpha_-)$ and Taylor expanding we
find
\begin{equation}
\tau(\alpha_-) = \frac{n\, log(z)}{2\pi i} + \cdots,
\end{equation}
which induces a monodromy $\tau \mapsto \tau+n$ upon encircling
$z=0$. This is the expected monodromy of a stack of $n$ D7-branes.
The other solution $\tau(\alpha_+)$ is S-dual to
$\tau(\alpha_-)$.

Now consider $n\geq 4$ D7-branes that are on top of an O7-plane, which in F-theory language corresponds to an $I_{n-4}^*$ fiber.
In this case the Weierstrass model is
\begin{equation}
f = z^2 F, \qquad g = z^3 G, \qquad \Delta = z^{2+n} \tilde \Delta,
\end{equation}
with $J$-invariant
\begin{equation}
J(I_{n-4}^*) = \frac{4F^3}{z^{n-4}\tilde \Delta}=: \frac{C}{z^{n-4}}.
\end{equation}
Then again solving the theory where $J$ is a quadratic in $x_1$ we obtain
(with similar $\alpha_\pm$)
\begin{equation}
\tau(\alpha_-) = \frac{(n-4)\, log(z)}{2\pi i} + \cdots,
\end{equation}
and there is a monodromy $\tau \mapsto \tau + n-4$ upon encircling $z=0$.
This is the monodromy expected for $n$ D7-branes on top of an O7-plane, and 
famously there is no monodromy in the case $n=4$, since the 4 D7-branes
cancel the charge of the O7-plane.

\subsection{Axiodilaton Profiles Near Seven-Branes with $J=1$}

From Tables \ref{tab:fibs} and \ref{tab:Jw} we see that the seven-branes with 
$J=1$ have fiber of Kodaira type $III$ and $III^*$, which carry gauge
symmetry $SU(2)$ and $E_7$, respectively. In both cases we have the same local
structure of the $J$-invariant
\begin{equation}
  J(III) = J(III^*) = \frac{1}{1+Az}.
\end{equation}
Using equation (\ref{eq:xinJquad}) we see 
\begin{equation}
x_1 = \frac{1\pm i \sqrt{A z}}{2} =: \alpha_\pm,
\end{equation}
which exhibits a $\bZ_2$ monodromy around $z=0$ that induces a monodromy on $\tau$. 
Using the relationship \eqref{eq:altq} between $q^{2\pi i \tau}$ and $x_1$ we compute two
values for $\tau$
\begin{equation}
\tau_{\pm} = i \, \frac{_2F_1(\frac16,\frac56,1,\alpha_+)}{_2F_1(\frac16,\frac56,1,\alpha_-)}.
\label{eq:tauexactJis1}
\end{equation}
Since the $\bZ_2$ monodromy swaps $\alpha_\pm$, it also swaps $\tau_\pm$ and noting $\tau_- = -1/\tau_+$
we see
\begin{equation}
\tau_\pm \mapsto \tau_\mp= - \frac{1}{\tau_\pm}
\end{equation}
under the monodromy. This matches the behavior associated with the monodromy matrices
\begin{equation}
M_{III}\begin{pmatrix} 0 & 1 \\ -1 & 0\end{pmatrix}, \qquad M_{III^*} = \begin{pmatrix} 0 & -1 \\ 1 & 0\end{pmatrix} 
\end{equation}
and we have seen the result by explicitly solving for the axiodilaton $\tau$. Note that this monodromy
is precisely an $S$-duality, which therefore also swaps electrons and monopoles represented by $(p,q)$-strings
to D3-brane probes.

How does the physics, in particular the string coupling, change upon moving away from the seven-brane? Expanding
the exact solution (\ref{eq:x2nearJis1}) near $z=0$ we obtain
\begin{equation}
\tau_\pm = i \pm B\sqrt{Az}-\frac{i}{2} B^2 A z + O(z^{3/2})
\end{equation}
where 
\begin{equation}
B = \frac{5\, \Gamma(\frac{7}{12}) \Gamma(\frac{11}{12})}{36\, \Gamma(\frac{13}{12}) \Gamma(\frac{17}{12})} \simeq .2638,
\end{equation}
is a constant that depends on values of the Euler $\Gamma$ function but $A = 27G^2/4F^3$ depends on the location
in the base. We see that $\tau(z)$ satisfies $\tau(0)=i$ and 
\begin{equation}
g_{s,\pm} \simeq \frac{1}{1 \pm B \,Im(\sqrt{Az}) - \frac12 B^2 Re(Az)}. 
\end{equation}
Close to $z=0$ we have
\begin{equation}
g_{s,\pm} \simeq \frac{1}{1 \pm B \,Im(\sqrt{Az})}, 
\end{equation}
and we see that the monodromy exchanges a more weakly coupled theory with a more strongly coupled theory, where the deviation from
$g_s=1$ depends on the model-dependent factor $A$ and the separation $z$ from the brane. This was also implicit from $\tau\mapsto -1/\tau$.

We see directly that the string coupling $g_s\simeq O(1)$ in the
vicinity of the type $III$ and type $III^*$ seven-branes carrying
$SU(2)$ and $E_7$ gauge symmetry, respectively, and that Ramanujan's
theory where $J$ is a quadratic in $x_1$ gives a set of $\tau$ values
permuted by the brane-sourced monodromy. From (\ref{eq:cantgoweak}),
an $SL(2,\bZ)$ transformation cannot make the theory weakly coupled in
this region.  \vspace{.5cm}

In the
previous section we saw that this method is more natural than the
theory where $J$ is cubic in $x_2$ since the $SL(2,\bZ)$ monodromy of
the type $III$ and $III^*$ Kodaira fibers is $\bZ_2$ rather than
$\bZ_3$. For completeness, though, inverting using the cubic theory gives
three solutions for $x_2$
\begin{align}
x_2 = \left\{-\frac{3}{A z}-\frac{11}{9}+O\left(z\right),\frac{1}{9}-\frac{8 i \sqrt{A^3} \sqrt{z}}{27 \sqrt{3}
   A}+O\left(z\right),\frac{1}{9}+\frac{8 i \sqrt{A^3}
   \sqrt{z}}{27 \sqrt{3} A}+O\left(z\right)\right\},
\end{align}
where we see that the first solution decouples near $z=0$ and one is left with the latter two
solutions, which we will call $\beta_\pm$. There is a $\bZ_2$ monodromy that exchanges $\beta_\pm$
upon encircling $z=0$. Defining $\tau_\pm$ for the cubic theory as 
$\tau_\pm \equiv \tau(\beta_\pm)$ we have
\begin{equation}
\tau_\pm = \frac{i}{\sqrt{2}} \frac{_2F_1(\frac14,\frac34,1,1-\beta_\pm)}{_2F_1(\frac14,\frac34,1,\beta_\pm)},
\end{equation}
and, though direct evaluation gives $\tau_\pm=i$ at $z=0$, the monodromy
$\tau \mapsto -\frac{1}{\tau}$ is not immediate, instead requiring the use of some identities for the
hypergeometric function for an exact expression.
We have verified via Taylor expansion that $\tau_\pm$ are swapped by a $\bZ_2$ monodromy, however.
Thus, the theory quadratic in $x_1$ seems better suited to study $III$ and $III^*$ fibers.

\subsection{Axiodilaton Profiles Near Seven-Branes with $J=0$}

We now turn to the study of seven-branes with Kodaira fibers satisfying $J=0$. 
From Tables \ref{tab:fibs} and \ref{tab:Jw} we see that the seven-branes with 
$J=0$ have general fiber of Kodaira type $II$, $II^*$, $IV$, and $IV^*$. The
seven-branes of the first two types carry no geometric gauge symmetry and $E_8$,
respectively, whereas the latter two exhibit $SU(3)$ ($SU(2)$) and $E_6$ ($F_4$)
geometric gauge symmetry respectively, if the geometry does not (does) exhibit
outer-monodromy that reduces the rank of the gauge group. Recall that
\begin{align}
  J(II) = J(IV^*) = \frac{z}{A+z}, \qquad J(II^*) = J(IV) = \frac{z^2}{A+z^2},
\label{eq:localJforJis1}
\end{align}
and that there is no discrepancy in this matching because, though the $SL(2,\bZ)$ monodromy
associated with these fibers are either order $3$ or $6$,  they are only order
 $3$ on $\tau$.

Let us utilize the theory where $J$ is a cubic in $x_2$ to study $\tau$ near these
seven-branes, beginning with the cases of a type $II$ and $IV^*$ fiber since these
have the same local structure of $J$-invariant. The expansion of the three solutions
to the cubic in $x_2$ expanded around $z=0$ are
\begin{equation}
x_2 =  -\frac13 - \frac{2(2A^2)^{1/3}}{3A} e^{\frac{2\pi i n}{3}} \, z^{1/3} -\frac{4}{3(2A^2)^{1/3}} e^{\frac{4\pi i n}{3}} \, z^{2/3} - \frac{1}{A} z + O(z^{4/3}), \qquad n\in\{0,1,2\},
\end{equation}
from which we see a $\bZ_3$ monodromy upon encircling the seven-brane
at $z=0$. Letting $\beta_n$ be the $x_2$ solution for each $n\in
\{0,1,2\}$, we have
\begin{equation}
\tau_n := \tau(\beta_n) = \frac{i}{\sqrt{2}} \frac{_2F_1(\frac14, \frac34, 1,1-\beta_n)}{_2F_1(\frac14, \frac34, 1,\beta_n)}.
\end{equation}
If the $\tau$ values are distinct then there is an order $3$ $SL(2,\bZ)$ monodromy on $\tau$, but its determining that
its precise action is $\tau \mapsto \frac{\tau-1}{\tau}$
would require using some identities of hypergeometric functions, unlike in the case of the type $III$ and $III^*$ seven-branes
where its action $\tau\mapsto -1/\tau$ was immediately clear. Instead we will prove the monodromy numerically in a power series in $z$.
Numerically at leading order in $z$ and keeping four significant figures in $z^{1/3}$, we have
\begin{align}
\tau_0 &\simeq e^{\frac{\pi i}{3}} -\frac{.3355i}{A^{2/3}} \, z^{1/3} + O(z^{2/3}) \nonumber \\
\tau_1 &\simeq e^{\frac{\pi i}{3}} +\frac{.2906+.1678 i}{A^{2/3}} \, z^{1/3} + O(z^{2/3})  \nonumber \\
\tau_2 &\simeq e^{\frac{\pi i}{3}} -\frac{.2906-.1678i }{A^{2/3}} \, z^{1/3} + O(z^{2/3}).
\label{eq:tau012forIIIVs}
\end{align}
We encircle $z=0$ by writing $z=re^{i\theta}$ where $r\in \bR$ is a
small positive number and then varying $\theta$.  There is a choice of
direction: encircling by taking $\theta$ from $0$ to $2\pi$ we see
that $\tau_0 \mapsto \tau_1$, $\tau_1\mapsto \tau_2$,
$\tau_2\mapsto\tau_0$, whereas going in the other direction by taking
$\theta$ from $0$ to $-2\pi$ gives the inverse action $\tau_0 \mapsto
\tau_2$, $\tau_1\mapsto \tau_0$, $\tau_2\mapsto\tau_1$. One can verify that
this latter action corresponds to the monodromy $\tau \mapsto \frac{\tau-1}{\tau}$;
i.e. $\frac{\tau_i - 1}{\tau_i} = \tau_{i-1}$ where $\tau_{-1}:=\tau_2$.

Now consider the cases of seven-branes with a
type $II^*$ and type $IV$ fiber. From (\ref{eq:localJforJis1}) we see that this differs 
from the analysis we just performed by the replacement $z\mapsto z^2$, as can be verified
by direct computation. The solutions to the cubic are
\begin{equation}
x_2 =  -\frac13 - \frac{2(2A^2)^{1/3}}{3A} e^{\frac{2\pi i n}{3}} \, z^{2/3} -\frac{4}{3(2A^2)^{1/3}} e^{\frac{4\pi i n}{3}} \, z^{4/3} - \frac{1}{A} z^2 + O(z^{8/3}), \qquad n\in\{0,1,2\},
\end{equation}
defining $\beta_n$ to be these three solutions the function form of $\tau$, and therefore $\tau_n:=\tau(\beta_n)$, remains the same.
Numerically at leading order in $z$, keeping four significant figures in $z^{2/3}$
we have 
\begin{align}
\tau_0 &\simeq e^{\frac{\pi i}{3}} -\frac{.3355i}{A^{2/3}} \, z^{2/3} + O(z^{4/3}) \nonumber \\
\tau_1 &\simeq e^{\frac{\pi i}{3}} +\frac{.2906+.1678 i}{A^{2/3}} \, z^{2/3} + O(z^{4/3})  \nonumber \\
\tau_2 &\simeq e^{\frac{\pi i}{3}} -\frac{.2906-.1678i }{A^{2/3}} \, z^{2/3} + O(z^{4/3}).
\label{eq:tau012forIIIVs}
\end{align}
Note that the monodromy action has changed, though: upon taking $\theta$ from $0$ to $2 \pi$
we have $\tau_0\mapsto \tau_2$, $\tau_1\mapsto \tau_0$, $\tau_2\mapsto \tau_1$. In the type $II$, $IV^*$
case this was the monodromy associated with taking $\theta$ from $0$ to $-2\pi$. We see explicitly
that the $SL(2,\bZ)$ monodromy of $II/IV^*$ fibers induce the inverse action on $\tau$ compared to
$II^*/IV$ fibers. This is as expected since
\begin{equation}
M_{II}=M_{II^*}^{-1}, \qquad M_{IV}=M_{IV^*}^{-1}
\end{equation}
and the fact
\begin{equation}
M_{II}=-M_{IV^*}. \qquad M_{II^*}=-M_{IV}
\end{equation}
implies that $II$ and $IV^*$ induce the same action on $\tau$, and similarly for $II^*$ and $IV$.

Given these explicit solutions for $\tau$ one can solve for $g_s$ as a function of the local coordinate
$z$ near the seven-branes with type $II$, $IV^*$, $II^*$, or $IV$ fibers and determine its falloff
from the central value $g_s=2/\sqrt{3}$ at $z=0$.

\subsection{The Genericity of Strongly Coupled Regions in F-theory and the Sen Limit}

In this section we would like to demonstrate the genericity of
strongly coupled regions in F-theory. This statement is immediately
plausible since F-theory is a generalization of the weakly coupled
type IIB string with varying axiodilaton, but we would like to argue
that there are regions with $O(1)$ $g_s$ for nearly all of the moduli
space of F-theory using two concrete lines of evidence, one that utilizes
non-Higgsable clusters and one that does not.

Recall from section \ref{sec:review} that there is growing evidence and argumentation
that nearly all extra dimensional topologies $B$ give rise to geometrically non-Higgsable
structure, and it is typically the case that those have geometrically
non-Higgsable seven-branes. The latter have a Kodaira fiber in the set
\begin{equation}\{II,III,IV,I_0^*,IV^*,III^*,II^*\}\end{equation} and
for all but\footnote{It would also be interesting to understand
  whether a non-Higgsable seven-brane with $I_0^*$ fiber, which could
  have gauge group $SO(8), SO(7),$ or $G_2$, necessarily forbids a Sen
  limit; certainly if it is Higgsable, such a limit can exist.}
$I_0^*$ the axiodilaton is strongly coupled in the vicinity of the
seven-brane, as seen using the explicit solutions of section
\ref{sec:axiodilaton}. If the argumentation regarding the genericity of non-Higgsable
clusters is correct and a typical compactification has a non-Higgsable
seven-brane with fiber $II,III,IV,IV^*, III^*,$ or $II^*$, then there are
regions with $O(1)$ $g_s$ for most of the moduli space of F-theory.

However, there are regions with $O(1)$ $g_s$
quite generally even in the absence of non-Higgsable clusters.
Consider a completely general Weierstrass model, which has a
$J$-invariant
\begin{equation}
J = \frac{4f^3}{4f^3+27g^2}.
\end{equation}
In some cases $f$ and $g$ may be reducible (e.g. if there are
non-Higgsable clusters), but it need not be so and it will not affect
the following calculation. There are two special loci in this generic
geometry, $f=0$ and $g=0$, and on these loci
\begin{equation}
J|_{f=0} = 0, \qquad J|_{g=0}=1.
\end{equation}
Since we have seen its utility for studying loci with $J=1$, consider Ramanujan's theory in
which $J$ is a quadratic in $x_1$. Then we have
\begin{equation}
x_1 = \frac12 \pm \frac{3i\sqrt{3} g}{4f^{3/2}}=: \alpha_\pm
\end{equation}
and solving for $\tau$ we have
\begin{equation}
\tau_{\pm} = i \, \frac{_2F_1(\frac16,\frac56,1,\alpha_+)}{_2F_1(\frac16,\frac56,1,\alpha_-)}.
\label{eq:tauonfis0}
\end{equation}
which is the same form as for seven-branes with type $III$ or $III^*$ Kodaira fibers
as in \eqref{eq:tauexactJis1}, though $\alpha_\pm$ have different forms which critically
change the physics. For example, the change in $\alpha_\pm$ changes the local structure of $\tau$ to
\begin{equation}
\tau_\pm = i \pm B \frac{g}{f^{3/2}}+O(g^2)
\end{equation} where here $B = \frac{5\, \Gamma(\frac{7}{12}) \Gamma(\frac{11}{12})}{8\sqrt{3}\, \Gamma(\frac{13}{12}) \Gamma(\frac{17}{12})} \simeq .68542$, and we see (at leading order\footnote{This is
  extended to all orders by the fact that the power series expansion for $_2F_1(a,b;c;x)$ holds for $|x|<1$, which occurs for sufficiently
  small $g$ away from $f=0$.}) that there is no monodromy associated with taking a loop around $g=0$ for $f\ne0$.
This makes physical sense, because such a locus has no seven-branes, and therefore no source for $\tau$-monodromy! 
Nevertheless, the region $g=0$ is strongly coupled: at $g=0$, $\alpha_+=\alpha_-$, $\tau=i$, and therefore
$g_s=1$. Similar results can be obtained using the $x_2$ theory to solve for $\tau$ near $f=0$,
and in that case $x_2$ has three solutions that give $\tau=e^{\pi i/3}$, and therefore $g_s=\frac{2}{\sqrt{3}}$ at $f=0$.

These facts about strong coupling are quite general, independent of
the existence or non-existence of any non-trivial seven-brane
structure (i.e. with fiber other than $I_1$).  They hold even for
completely smooth models, such as the generic Weierstrass model over
$\bP^3$.

\vspace{.5cm}
What happens to these strongly coupled regions in the Sen limit?
Roughly, the Sen limit \cite{Sen:1996vd,Sen:1997gv} is a weakly
coupled limit in moduli space where $J\mapsto \infty$ and therefore
$\tau \mapsto i \infty$, so $g_s\mapsto 0$. This occurs because the
Weierstrass model takes the form $f=C \eta -3 h^2$, $g=h \left(C \eta -2 h^2\right)$
with discriminant
\begin{equation}
\Delta = C^2 \eta^2(4C\eta - 9h^2)
\end{equation}
where $\eta$ and $h$ are sections and $C$ is a parameter. Sen's weak coupling limit is the $C\mapsto 0$
limit. However, note that if $C$ is very small but non-zero
\begin{equation}
J = \frac{4f^3}{C^2 \eta^2(4C\eta - 9h^2)} = \frac{4 \left(C \eta -3 h^2\right)^3}{C^2 \eta ^2 \left(4 C \eta -9 h^2\right)}
\end{equation}
is made very large by $1/C$, but not infinite. Therefore unless it is strictly true that $C=0$,
the loci $f=0$ and $g=0$ (i.e. $C\eta=3h^2$ and $\{h=0\}\cup\{C\eta=2h^2\}$) have $J=0$ and $J=1$, and therefore
$g_s=2/\sqrt{3}$ and $g_s=1$, respectively. In the limit of $C$ becoming very small but non-zero one expects the string coupling to become lower 
in the vicinity of $f=0$ and $g=0$, though $O(1)$ on the loci. 

Let us see this explicitly. First, solving quadratic theory for $\tau$ near $h=0$ (a component of $g=0$) we obtain
\begin{equation}
\tau_\pm = i \pm B\, (C\eta)^{-1/2} \, h+O(h^2)
\end{equation}
where $B = \frac{5\, \Gamma(\frac{7}{12}) \Gamma(\frac{11}{12})}{8\sqrt{3}\, \Gamma(\frac{13}{12}) \Gamma(\frac{17}{12})} \simeq .68542$.
As $C$ becomes large, the $g_s$ associated with one of these solutions for a small $h$ becomes weaker, but $g_s=1$
at $h=0$ if $C$ is finite. Studying the other component of $g=0$ where $C\eta=2h^2$ is more difficult because the locus
itself moves as $C$ is taken to $0$. This component intersects a disc centered at $h=0$ at $h=\pm \sqrt{C\eta/2}$,
and solving the quadratic theory near $h=\sqrt{C\eta/2}$ we have
\begin{equation}
\tau_\pm = i \pm i\,4\sqrt{2}B\,\left(h-\sqrt{\frac{C\eta}{2}}\right)+O\left(\left[h-\sqrt{\frac{C\eta}{2}}\right]^2\right).
\end{equation}
In the strict limit $C=0$ all of these loci collapse to $h=0$ and the theory is weakly
coupled, but any $C\ne 0$ gives $g_s=1$ on the two
components of $g=0$. Similarly, the locus $f=0$ gives rise to $g_s=2/\sqrt{3}$ 
point at the loci $h=\pm \sqrt{\frac{C\eta}{3}}$ in an $h$-disc, and this could
be studied explicitly using the theory where $J$ is cubic in $x_2$.

What is going on physically for these components? We have three: one for $f=0$
and two for $g=0$, and unless $g_s=0$ they are strongly coupled loci.
Examining the discriminant we see that as $C$ becomes small two
seven-branes approach the locus $h=0$ and collide in the limit. This
is the $O7$-plane, and therefore for $C\ne 0$ the region $h=0$ is the
strongly coupled region between the two seven-branes that become the
$O7$ in the Sen limit. The $h$-disc is useful for studying $f=0$ and the
other component of $g=0$. The picture is
\begin{equation} 
\begin{tikzpicture}[scale=1]
    \draw[xshift=7cm,thick,color=black] (60mm,0mm) circle (1mm);
    \draw[xshift=7cm,thick,color=black] (-60mm,0mm) circle (1mm);
    \fill[xshift=7cm,thick,color=blue] (0mm,0mm) circle (1mm);
    \fill[xshift=7cm,thick,color=blue] (28.2843mm,0mm) circle (1mm);
    \fill[xshift=7cm,thick,color=blue] (-28.2843mm,0mm) circle (1mm);
    \fill[xshift=7cm,thick,color=red] (23.094mm,0mm) circle (1mm);
    \fill[xshift=7cm,thick,color=red] (-23.094mm,0mm) circle (1mm);
\end{tikzpicture}
\label{eqn:pixpiypiz}
\end{equation}
where the hollowed circles are the only places where $\Delta$
intersects the $h$-disc. These are the two $(p,q)$ seven-branes that
become the $O7$ in the Sen limit, and in this limit the whole pictures
collapse to the central blue dot at $h=0$. The red (blue) dots are
where the $f=0$ ($g=0$) locus intersects the $h$-disc, and they have
$g_s=2/\sqrt{3}$ and $g_s=1$, respectively. They are strongly coupled
regions that are separated from the branes for finite $C$.

\vs

The one caveat that we have not yet discussed involves configurations with
constant coupling. They were studied in the K3 case by Dasgupta and
Mukhi \cite{Dasgupta:1996ij}, with the result (which generalizes beyond
K3) that seven-branes with fiber $II$, $III$, $IV$, $I_0^*$, $IV^*$,
$III^*$, and $II^*$ may gave rise to constant coupling
configurations. When this occurs, all but the $I_0^*$ case gives
rise to constant coupling configurations with $O(1)$ $g_s$. In the
$I_0^*$ case there are a continuum of possible $g_s$ values that may
be constant cross the base $B$. In such a case there may be multiple $I_0^*$ seven-branes
and the Weierstrass model
takes the form
\begin{equation}
f = F \prod_i z_i^2, \qquad g = G \prod_i z_i^3,
\end{equation}
with $F$ and $G$ necessarily constants, rather than non-trivial sections of a line bundle. In such a case
they do not define vanishing loci in $B$ with $O(1)$ $g_s$. Instead, all of the factors of $z_i$ drop out
of the $J$ invariant,
\begin{equation}
J = \frac{4F^3}{4F^3+27G^2}
\end{equation}
which is just a constant, not varying over the base. To our knowledge,
this is the only possible way to obtain an F-theory compatification
with non-zero $g_s\ll 1$ which is weakly coupled everywhere in $B$.

In summary, aside from compactifications realizing this $I_0^*$ caveat
or the strict $g_s=0$ limit, there is a region $f=0$ or $g=0$ (or a component thereof) in every
F-theory compactification with $O(1)$ $g_s$, and it is not necessarily near
any seven-brane.  This may have interesting implications for moduli
stabilization or cosmology in the landscape.

\section{Non-Perturbative $SU(3)$ and $SU(2)$ Theories}
\label{sec:su3su2}

\subsection{Comparison to D7-brane Theories}

In this section we would like to compare the non-perturbative
realizations of $SU(3)$ and $SU(2)$ theories from type $IV$ and $III$
fibers to the $SU(3)$ and $SU(2)$ theories\footnote{The type $IV$ and $I_3$ fibers
realize $SU(2)$ ($SU(3)$) theories in six and four-dimensional models if the 
fiber is non-split (split).} that may be realized by
stacks of three and two $D7$-branes at weak string coupling. The
latter are realized by type $I_3$ and $I_2$ fibers, and in some cases
(but not if the $IV$ or $III$ is non-Higgsable) they are related by
deformation.

Such deformations are slightly unusual. In many cases a deformation of
a geometry spontaneously breaks the theory on the seven-brane, but in
these cases the deformation leaves the gauge group
intact\footnote{This is always true for a $III$ to $I_2$ deformation,
  and is also true for a $IV$ to $I_3$ deformation provided the deformation preserve
  the split or non-split property related to non-simply laced groups.}. However, the deformation is
non-trivial, since the Kodaira fiber, the number of branes in the stack, the $SL(2,\bZ)$ monodromy,
and the axiodilaton profile all change due to the deformation.

We begin by considering the relationship between $SU(2)$ theories
realized on seven-branes with type $I_2$ and $III$ Kodaira fibers.
We expand $f$ and $g$ as
\begin{equation}
f = f_0 + f_1 z \qquad \qquad g = g_0 + g_1 z + g_2 z^2,
\end{equation}
where $f_0$, $g_0$, and $g_1$ do not depend on $z$ but $f_1$ and $g_2$
can contain terms that are both constant in $z$ and depend on $z$. To engineer
an $I_2$ singularity we move to a sublocus in complex structure moduli where
\begin{equation}
f_0 = -3a^2 \qquad g_0 = 2a^3 \qquad f_1 = b \qquad g_1 = -a b,
\end{equation}
in which case
\begin{equation}
\Delta = z^2 \left(108 a^3 g_2-9 a^2 b^2\right)+ z^3(4b^3-54abg_2)+O(z^4).
\end{equation}
Here $a$ and $b$ are global sections of line bundles $a\in
\Gamma(\cO(-2K))$ and $b \in \Gamma(\cO(-4K-Z))$ where $Z$ is the class of the divisor
$z=0$. Specifying in moduli such that $a=0$, we see that $f_0$, $g_0$, and $g_1$ vanish
identically, and the resulting fibration has $(ord f, ord g, ord \Delta) = (1,2,3)$; thus
in the limit $a\mapsto 0$ the $I_2$ fiber over $z=0$ becomes type $III$.

In a generic disc $D$ containing $z=0$ where $z$ is also the coordinate of
the disc, $a$ is a constant since $f_0$, $g_0$, and $g_1$ 
do not depend on $z$. We can study the geometry
of the elliptic surface over the disc that is naturally induced from
restriction of the elliptic fibration. From the point of view of the
elliptic surface, then, the $a\mapsto 0$ limit is just a limit in a
constant complex number $a$ (technically $a|_D$, but we abuse
notation). Then $a=0$ realizes a type $III$ fiber along $z=0$, and a
small deformation $a\ne 0$ reduces it to an $I_2$ fiber; both
geometrically give rise to $SU(2)$ gauge theories on the
seven-brane.

 We would also like to consider a second deformation
parameter $\epsilon$, set $b=g_2=1$ for simplicity, and consider a small
enough disc that we can drop higher order terms in $z$; then the
two-parameter family of elliptic fibrations over $D$ is given by
\begin{align}
f = -3a^2 + z \qquad \qquad g = \epsilon + 2 a^3 - a z + z^2  \nonumber \\
\Delta = 108 a^3 z^2+108 a^3 \epsilon -9 a^2 z^2-54 a z^3-54 a z \epsilon +27 z^4+4 z^3+54 z^2
   \epsilon +27 \epsilon ^2.
\end{align}
We note the behavior of the discriminant in the relevant limits:
\begin{align}
\epsilon=0&:  \Delta = z^2 \left(108 a^3-9 a^2-54 a z+27 z^2+4 z\right)\\
\epsilon=a=0&: \Delta = z^3 \left(27 z+4\right),
\end{align}
where the $\epsilon=0$ limit is the limit of $SU(2)$ gauge enhancement with a type
$I_2$ fiber, and the $\epsilon=a=0$ limit maintains the $SU(2)$ group but realizes
the theory instead with a type $III$ fiber.

\vspace{.5cm}
\noindent \emph{Analysis of the Two Deformations}

\begin{figure}[t]
\begin{center}
\includegraphics[scale=.75]{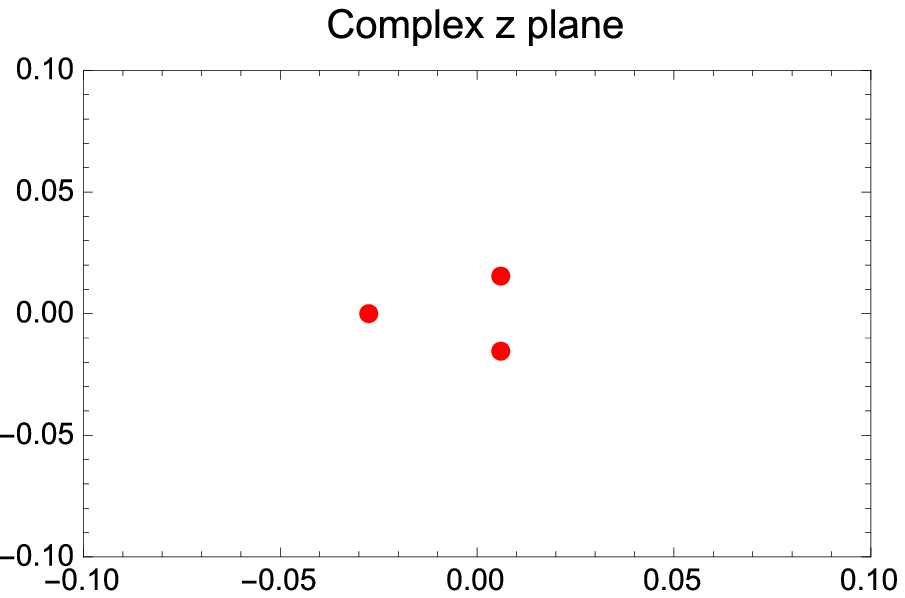} \qquad \qquad
\includegraphics[scale=.75]{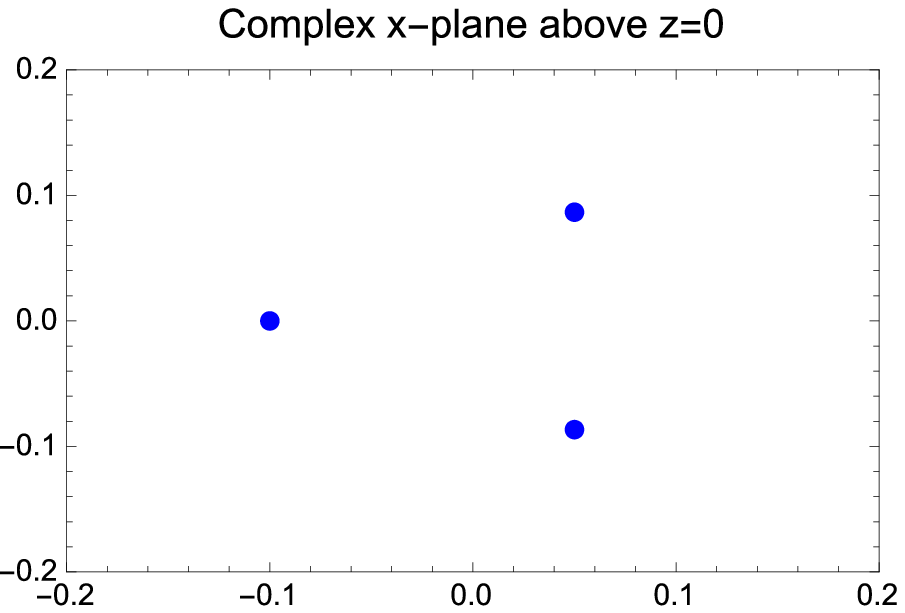}
\end{center}
\caption{Two figures of the geometry deformed by $\epsilon=.001$, $a=0$, with
$(p,q)$ seven-branes on the left and ramification points of the elliptic curve on the right.}
\label{fig:a0}
\end{figure}

Let us study two different deformations of the type $III$
theory that uncover its differences from the $I_2$ theory, and also
the relationship between the two via deformation. The first
deformation is to take $\epsilon \ne 0$ but small; specifically,
$\epsilon = .001$. This is a small breaking of the type $III$ $SU(2)$
theory, where the deformation causes the three branes comprising the
type $III$ theory to split into three $(p,q)$ seven-branes in a smooth
geometry with no non-abelian gauge symmetry. For these parameters the
geometry appears in Figure \ref{fig:a0}
where we have displayed the branes in the $z$-plane as red dots and we
have also displayed the $x$-plane above $z=0$, where the blue dots
represent three ramificaton points of the torus described as a double
cover (as is natural in a Weierstrass model). Any straight line
between two of the blue dots determines a one-cycle in the elliptic
fiber above $z=0$, and by following straight line paths from $z=0$ to
the seven-branes two of the ramification points will collide,
determining a vanishing cycle.

Obtaining a consistent picture of the W-boson degrees of freedom throughout
the moduli space requires that the geometry provides a mechanism that turns the
massive W-bosons of the slightly deformed type $III$ theory to the massive W-bosons 
of the slightly deformed type $I_2$ theory. From the weak coupling limit, we know that
the latter are represented by fundamental open strings. From the deformation of the
type $III$ singularity performed in \cite{Grassi:2014sda} we know that the former are 
three-pronged string junctions. Thus, for $\epsilon\ne 0$ small, the continual
increase of $a$ from $0$ must turn string junctions into fundamental strings.

For the deformation of a type $III$ singularity the vanishing cycles were
derived in \cite{Grassi:2014sda}; they can be read off by taking
straight line paths from the origin. Beginning with the left-most brane
and working clockwise about $z=0$, the ordered set of vanishing cycles are $Z_{III}=\{\pi_2,\pi_1,\pi_3\}$
where these cycles are defined as
\begin{equation} 
\begin{tikzpicture}[scale=1]
    \fill[xshift=7cm,thick] (180:10mm) circle (1mm);
    \fill[xshift=7cm,thick] (180-120:10mm) circle (1mm);
    \fill[xshift=7cm,thick] (180+120:10mm) circle (1mm);
    \node at (9.2cm,1.3cm) {$x$};
    \draw[xshift=7cm,thick,->] (180:10mm)+(30:1.3mm) -- +(30:16mm);
    \draw[xshift=7cm,thick,->] (180-120:10mm)+(-90:1.3mm) -- +(-90:16mm);
    \draw[xshift=7cm,thick,->] (180+120:10mm)+(150:1.3mm) -- +(150:16mm);
    \node at (6.6cm,0.7cm) {$\pi_1$};
    \node at (8cm,0cm) {$\pi_2$};
    \node at (6.6cm,-0.7cm) {$\pi_3$};
    \draw[xshift=9cm,thick,yshift=1.0cm] (90:0mm) -- (90:4mm);
    \draw[xshift=9cm,thick,yshift=1.0cm] (0:0mm) -- (0:4mm);
 \end{tikzpicture}
\label{eqn:pixpiypiz}
\end{equation}
and the massive W-bosons of the broken theory are three-pronged string
junctions $J_\pm=(\pm 1,\pm 1,\pm 1)$ which, topologically, are two
spheres in the elliptic surface due to having asymptotic  charge zero \cite{Grassi:2014ffa}.
Pictorially, they appear as
\begin{equation}
  \begin{tikzpicture}
  \begin{scope}[rotate=-30]  
  \fill[xshift=-20mm] (90:8mm) circle (1mm);
  \fill[xshift=-20mm] (210:8mm) circle (1mm);
  \fill[xshift=-20mm] (330:8mm) circle (1mm);
  \draw[xshift=-20mm,thick] (0:0mm) -- (90:7mm);
  \draw[xshift=-20mm,thick] (90:3.5mm) -- (75:3mm);
  \draw[xshift=-20mm,thick] (90:3.5mm) -- (105:3mm);
  \draw[xshift=-20mm,thick] (0:0mm) -- (210:7mm);
  \draw[xshift=-20mm,thick] (210:3.5mm) -- (225:3mm);
  \draw[xshift=-20mm,thick] (210:3.5mm) -- (195:3mm);
  \draw[xshift=-20mm,thick] (0:0mm) -- (330:7mm); 
  \draw[xshift=-20mm,thick] (330:3.5mm) -- (345:3mm);
  \draw[xshift=-20mm,thick] (330:3.5mm) -- (315:3mm);
  \end{scope}
  \begin{scope}[yshift=2cm,rotate=-30]
  \fill[xshift=20mm] (90:8mm) circle (1mm);
  \fill[xshift=20mm] (210:8mm) circle (1mm);
  \fill[xshift=20mm] (330:8mm) circle (1mm);
  \draw[xshift=20mm,thick] (90:0mm) -- (90:7mm);
  \draw[xshift=20mm,thick] (90:3.5mm) -- (78:4.1mm);
  \draw[xshift=20mm,thick] (90:3.5mm) -- (102:4.1mm);
  \draw[xshift=20mm,thick] (0:0mm) -- (210:7mm);
  \draw[xshift=20mm,thick] (210:3.5mm) -- (222:4.1mm);
  \draw[xshift=20mm,thick] (210:3.5mm) -- (198:4.1mm);
  \draw[xshift=20mm,thick] (0:0mm) -- (330:7mm); 
  \draw[xshift=20mm,thick] (330:3.5mm) -- (342:4.1mm);
  \draw[xshift=20mm,thick] (330:3.5mm) -- (318:4.1mm);
  \end{scope}
  \end{tikzpicture}
\end{equation}
where the black dots represent seven-branes. See \cite{Grassi:2014sda} for more details of the intersection
theory of these particular junctions, as well as their reproduction of the
Dynkin diagram.

\vspace{.3cm} The first deformation should be thought of as a small
deformation of an $SU(2)$ seven-brane theory associated with a type $III$
singularity, that is, a Higgsing.  We would now like to study a deformation that
corresponds to a large deformation from a type $III$ $SU(2)$ theory to a 
type $I_2$ $SU(2)$ theory (which does not Higgs the theory), and then a small
deformation that Higgses the $SU(2)$ theory on the seven-brane with an $I_2$
singular fiber. Heuristically, the third brane of the type $III$ singularity 
should be far away relative to the distance between the two branes of the deformed
type $I_2$ singularity, which in type IIB language are two D7-branes with a small
splitting.

\begin{figure}[t]
\begin{center}
\includegraphics[scale=.75]{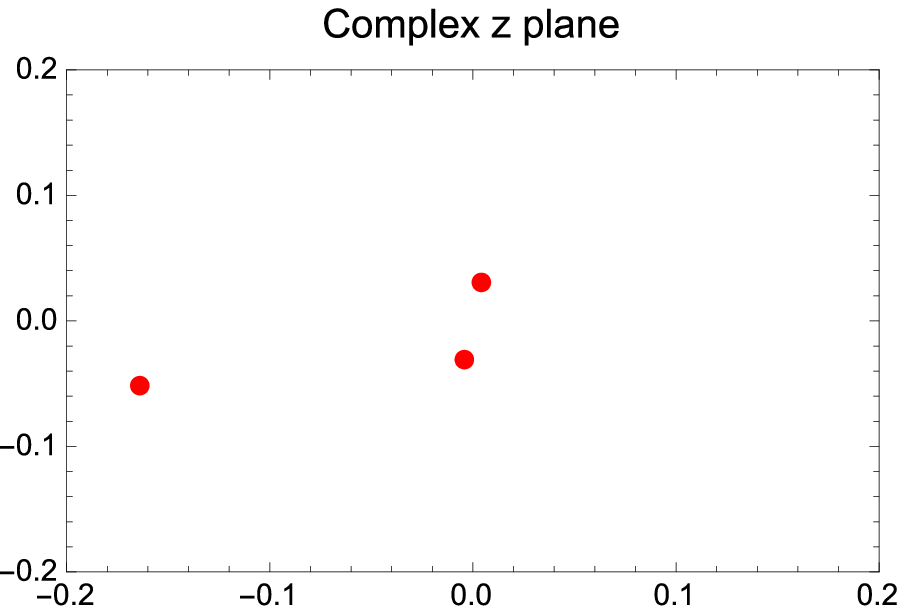} \qquad \qquad
\includegraphics[scale=.75]{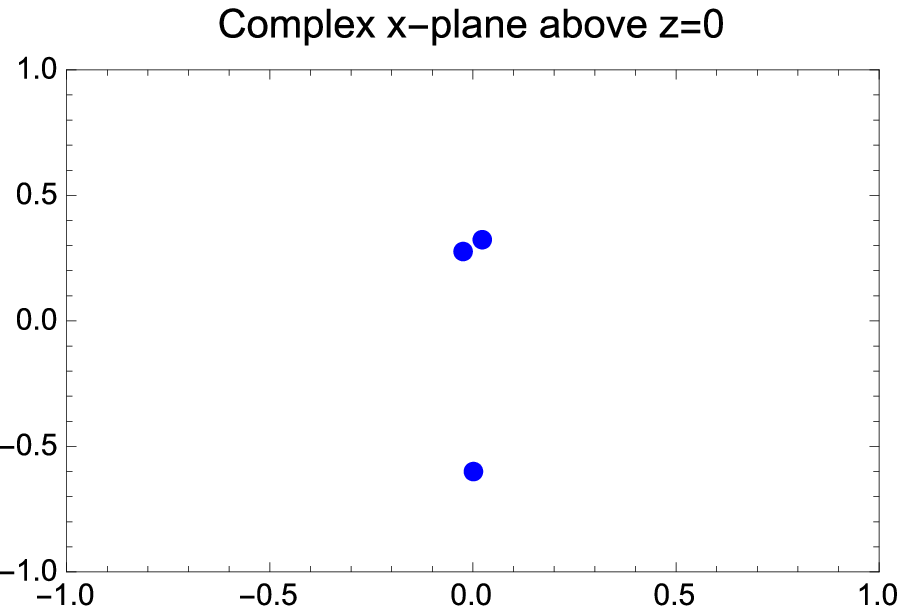}
\end{center}
\caption{Two figures of the geometry deformed by $\epsilon=.001$, $a=.3i$, with
$(p,q)$ seven-branes on the left and ramification points of the elliptic curve on the right.}
\label{fig:an0}
\end{figure}

We take $a=.3i$ and $\epsilon=.001$. The difference between
this deformation and the deformation of the previous paragraph is that
in this case one of the three seven-branes that was originally at $z=0$ is
much further away due to taking $a\ne 0$, and the question is whether
the $(p,q)$ labels of the three-branes changed in the process. The geometry
appears in Figure \ref{fig:an0}
where the brane on the left is the one that has been moved further away
from the origin by turning on $a$. If one were to maintain this value of $a$
but tune $\epsilon\mapsto 0$, the two branes close to the origin would collide
to give an  $SU(2)$ theory with an $I_2$ singular fiber.
Indeed, with this deformation the two-branes closer
to the origin both have vanishing cycle $\pi_1$ and the brane
displaced by the $a$ deformation has vanishing cycle $\pi_3$, so that
now the ordered set of vanishing cycles (beginning with the left-most brane and working
clockwise about $z=0$) is $Z=\{\pi_3,\pi_1,\pi_1\}$. The
W-bosons of the broken $I_2$ $SU(2)$ theory are the strings between the
$\pi_1$ branes, as expected since $I_2$ is the F-theory lift of two
D7-branes.

What has happened? The natural W-boson of the broken type $III$ SU(2)
theory (associated to the $a=0, \epsilon=.001$ deformation) is the
three-pronged string junction, but tuning $a$ from $a=0$ to $a=.3i$
the natural W-boson of the broken type $I_2$ theory is the fundamental
string. The change between the two different descriptions of the W-boson
is a deformation induced Hanany-Witten move; for related ideas in a different
geometry, see \cite{Cvetic:2010rq}. This can actually be seen via
continuous deformation from $a=0$ to $a=.3i$, during which the leftmost (bottom) seven-brane
in Figure \ref{fig:a0} becomes the bottom (leftmost) seven-brane of Figure \ref{fig:an0}.
In the process the straight line path from $z=0$ to the (moving) leftmost seven-brane in Figure \ref{fig:a0}
is crossed via the movement of the bottom seven-brane in Figure \ref{fig:a0}, changing the $(p,q)$ labels
of the former. This matches the changes in the ordered sets of vanishing cycles.

Though we have explicitly seen the natural state change using a
deformation, let us check the possibilities using the usual algebraic
description of Zwiebach and Dewolfe \cite{DeWolfeZwiebach} where the
$(p,q)$ seven-branes are arranged in a line with branch cuts pointing
downward. Technically, the branch cuts represent the mentioned
straight line paths from the origin, the latter being the point whose
associated fiber $E_0$ appears in the relative homology $H_2(X_d,E_0)$
that defines the string junctions. $X_d$ is the elliptic surface 
of the disc. The
W-boson of the broken $SU(2)$ theory associated to a type $III$
singularity is
\begin{center}
\begin{tikzpicture}
  \draw[thick,dotted] (0cm,-.25cm) -- (0cm,-1.75cm);
  \draw (0cm,0cm) circle (.25cm);
  \node at (0cm,0cm) {$\pi_3$};
  \draw[thick,dotted] (1cm,-.25cm) -- (1cm,-1.75cm);
  \draw (1cm,0cm) circle (.25cm);
  \node at (1cm,0cm) {$\pi_2$};
  \draw[thick,dotted] (-1cm,-.25cm) -- (-1cm,-1.75cm);
  \draw (-1cm,0cm) circle (.25cm);
  \node at (-1cm,0cm) {$\pi_1$};
  \draw[thick] (-1cm,.25cm) arc (135:45:1.4cm);
  \draw[thick] (0cm,.25cm) -- (0cm,.65cm);
  \draw[thick] (-.8cm,.5cm) -- (-.65cm,.5cm);
  \draw[thick] (-.66cm,.36cm) -- (-.66cm,.5cm);
  \draw[thick] (.8cm,.5cm) -- (.65cm,.5cm);
  \draw[thick] (.66cm,.36cm) -- (.66cm,.5cm);
  \draw[thick] (-.1cm,.4cm) -- (0cm,.5cm);
  \draw[thick] (.1cm,.4cm) -- (0cm,.5cm);
\end{tikzpicture}
\end{center}
where for concreteness we choose a basis such that $\pi_1 = (1,0)$, $\pi_2=(0,1)$ and
$\pi_3=(-1,-1)$ with the convention that we cross branch cuts by moving to the right. Then
the monodromy matrix associated to a $(p,q)$ seven-brane is 
\begin{equation}
  \begin{pmatrix}
    1-pq && p^2 \\
    -q^2 && 1+pq
  \end{pmatrix}
  \label{eq:Mpq}
\end{equation}
and one can check that the monodromy of these three branes reproduces the monodromy of the
type $III$ singularity, as they must after deformation. We see
\begin{equation}
M_{\pi_2}M_{\pi_3}M_{\pi_1}=M_{III}=
\begin{pmatrix}
  0 && 1 \\ -1 && 0
\end{pmatrix},
\end{equation}
which is the expected behavior.

In this description, the continuous motion of the seven-branes
described above amounts to the $\pi_3$ brane crossing the branch cut
of the $\pi_2$ brane, changing its vanishing cycle to
$M_{\pi_3}^{-1}\begin{pmatrix}0\\1\end{pmatrix}=\begin{pmatrix}-1\\0\end{pmatrix}$
seven-brane; but this is just $-\pi_1$ and we are free to call it a
$\pi_1$ seven-brane instead if we reverse the sign of any junction
coming out of the brane. With this movement and relabeling, the above
junction becomes
\begin{center}
\begin{tikzpicture}
  \draw[thick,dotted] (0cm,-.25cm) -- (0cm,-1.75cm);
  \draw (0cm,0cm) circle (.25cm);
  \node at (0cm,0cm) {$\pi_1$};
  \draw[thick,dotted] (1cm,-.25cm) -- (1cm,-1.75cm);
  \draw (1cm,0cm) circle (.25cm);
  \node at (1cm,0cm) {$\pi_3$};
  \draw[thick,dotted] (-1cm,-.25cm) -- (-1cm,-1.75cm);
  \draw (-1cm,0cm) circle (.25cm);
  \node at (-1cm,0cm) {$\pi_1$};
  \draw[thick] (1cm,.25cm) -- (1cm,.76cm);
  \draw[thick] (.9cm,.4cm) -- (1cm,.5cm);
  \draw[thick] (1.1cm,.4cm) -- (1cm,.5cm);
  \draw[thick] (-.52cm,.61cm) -- (-.37cm,.59cm);
  \draw[thick] (-.39cm,.44cm) -- (-.37cm,.59cm);
  \draw[thick] (.65cm,-.5cm) -- (.56cm,-.58cm);
  \draw[thick] (.63cm,-.7cm) -- (.56cm,-.58cm);
  \draw[thick] (-1cm,.25cm) .. controls (0cm,1cm) and (2cm,1cm) .. (2cm,0cm) .. controls (2cm,-1cm) and (.2cm,-.75cm) .. (.15cm,-.2cm);
\end{tikzpicture}
\end{center}
and now the branes are in a position to do the relevant Hanany-Witten move. Prior to crossing the branch
cut from the right, the $(p,q)$ charge of the piece of string that ends on the middle brane (in this picture) is $\pi_1+\pi_3$, and after the branch cut it is $\pi_1$ due
to the monodromy action $M_{\pi_3}^{-1}$. If the monodromy action is replaced by a prong by pulling the string
through the brane (that is, performing a Hanany-Witten move), how many extra prongs $k$ are picked up on the right-most brane?
Charge conservation requires 
\begin{equation} \pi_1 + \pi_3 + k \,\pi_3 = \pi_1 \end{equation} 
and we see $k=-1$. That is, the Hanany-Witten move replaces the
portion of the string crossing the branch cut with a prong going into the $\pi_3$ brane. This cancels against the prong that is already there, leaving
\vs
\begin{center}
\begin{tikzpicture}
  \draw[thick,dotted] (0cm,-.25cm) -- (0cm,-1.75cm);
  \draw (0cm,0cm) circle (.25cm);
  \node at (0cm,0cm) {$\pi_1$};
  \draw[thick,dotted] (1cm,-.25cm) -- (1cm,-1.75cm);
  \draw (1cm,0cm) circle (.25cm);
  \node at (1cm,0cm) {$\pi_3$};
  \draw[thick,dotted] (-1cm,-.25cm) -- (-1cm,-1.75cm);
  \draw (-1cm,0cm) circle (.25cm);
  \node at (-1cm,0cm) {$\pi_1$};
  \draw[thick] (-.75cm,0cm) -- (-.25cm,0cm);
  \draw[thick] (-.55cm,.1cm) -- (-.45cm,0cm);
  \draw[thick] (-.55cm,-.1cm) -- (-.45cm,0cm);
\end{tikzpicture}
\end{center}
which, since we have chosen $\pi_1=(1,0)$, is just a fundamental string.
If we had chosen $\pi_1$ to be some other $(p,q)$ value this would be a $(p,q)$
string, but there is always a choice of $SL(2,\bZ)$ frame what would turn
it back into a fundamental string.

In summary, we see using algebraic techniques that the natural W-boson
of the Higgsed type $III$ $SU(2)$ theory, which is a three-pronged
string junction, is related to the the natural W-boson of the Higgsed
type $I_2$ $SU(2)$ theory. The relationship is a brane rearrangement
together with a Hanany-Witten move, which we have seen explicitly
above via two deformations of the elliptic surface with a type $III$
singular fiber.

\vs

The natural W-bosons of the type $I_3$
theory are (in an appropriate $SL(2,\bZ)$ duality frame) fundamental
strings and the natural W-bosons of the type $IV$ theory include
string junctions (see e.g. \cite{Grassi:2014sda} for an explicit analysis).  Via deformation they must be related to one
another, and given the result we have just seen it is natural to expect a brane rearrangement
and a Hanany-Witten move. This expectation is correct, though we will not
explicitly show it since the techniques are so similar to the previous case.

We would like to present the deformation for the interested reader, though. Take 
\begin{equation}
  f=-3 a^4+a b\, z+c\, z^2+f_3\, z^3\qquad \qquad
  g=2 a^6-a^3 b\, z+ \left(\frac{b^2}{12}-a^2 c\right)\, z^2+g_3 z^3
\end{equation}
where $a$, $b$, and $c$ are global section $a\in \cO(-K)$, $b\in
\cO(-3K-Z)$ and $c \in \cO(-4K-Z)$ with $Z$ is the class of $z=0$.
Here $f_3$ and $g_3$ contain constant terms in $z$ as well as higher
order terms.  The discriminant is
\begin{align}
  \Delta &= \frac{1}{16} z^3 \left(1728 a^8 f_3+1728 a^6 g_3-288 a^5 b c-8 a^3 b^3\right)\nonumber \\ & +\frac{1}{16} z^4 \left(-1152 a^5
   b f_3-144 a^4 c^2-864 a^3 b g_3+120 a^2 b^2 c+3 b^4\right)+O(z^5)
\end{align}
and we see that there is a gauge theory with $I_3$ fiber on the seven-brane at $z=0$. In the $a=0$ limit we see that
\begin{align}
f = z^2 \left(c+f_3 z\right) \qquad \qquad g=\frac{1}{12} z^2 \left(b^2+12 g_3 z\right) \nonumber \\
\Delta = \frac{1}{16} z^4 \left(3 b^4+72 b^2 g_3 z+64 c^3 z^2+192 c^2 f_3 z^3+192 c
   f_3^2 z^4+64 f_3^3 z^5+432 g_3^2 z^2\right)
\end{align}
which has a gauge theory with type $IV$ Kodaira fiber on the seven-brane at $z=0$. In these cases we have not
imposed the absence of outer monodromy (i.e. we have no imposed a split fiber), so they are $SU(2)$ gauge theories. If the form is further restricted so
that outer monodromy is imposed, it is an $SU(3)$ gauge theory.

\newsec{Extra Branes at Seven-Brane Intersections}
\label{sec:extra branes}

Two seven-branes that intersect along a codimension two locus $C\subset B$ are typically only seven-branes that
intersect along $C$. However, a counterexample giving rise to $SU(3)\times SU(2)$ gauge symmetry 
was studied in \rcite{GrassiHalversonShanesonTaylor2014} where an additional brane with an $I_1$ singular fiber also intersected
the curve of $SU(3)\times SU(2)$ intersection. Though one might have expected a
$IV$-$III$ collision, a $IV$-$III$-$I_1$ collision occurs automatically. The Weierstrass model is
 $f= z^1 t^2\, F$ and $g=z^2 t^2\, G$ with
\eqn{\Delta = z^3 t^4\,(4t^2 F^3 + 27 z G^2 )=: z^3t^4\tilde \Delta,}
from which it can be seen that the brane along $\tilde \Delta = 0$ intersects $\{z=t=0\}=:C$. Three
stacks of branes intersect $C$, and we call $\tilde \Delta=0$ the ``extra brane'' at $C$.

There is a natural guess regarding the associated physics:
where there is an extra brane there should be extra string (or string junction) states.
In the mentioned example it was found that such a collision gives
rise not only to the expected bifundamentals of $SU(3)\times SU(2)$, but also \cite{GrassiHalversonShanesonTaylor2014}
fundamental hypermultiplets of $SU(3)$ and $SU(2)$ which could come in
chiral multiplets if flux is turned on. Such configurations may be non-Higgsable;
for example geometries, see \cite{GrassiHalversonShanesonTaylor2014,Halverson:2015jua,Taylor:2015ppa}.

\vs

Under which circumstances is there necessarily an extra brane?  We take $F_1$ and $F_2$ to be the
Kodaira singular fibers of seven-branes that collide along $C$, with 
\begin{equation}
  F_{1,2}\in\{II,III,IV,I_0^*,IV^*,III^*,II^*\}.\end{equation}
These are the possible fibers of non-Higgsable seven-branes.
Let
$J_1$ and $J_2$ be the $J$-invariants associated with $F_1$ and $F_2$. By direct calculation we find:
\begin{itemize}
	\item If either $F_1$ or $F_2$ is $I_0^*$ there is no additional brane.
	\item If neither $F_1$ nor $F_2$ are $I_0^*$ then there is an additional brane if and only if $J_1\ne J_2$.
	\item If there is an additional brane then precisely one of the fibers is type $III^*$ or $III$.
	\item If there is an additional brane and the Weierstrass
          model does not have $(f,g)$ vanishing to order $\geq (4,6)$
          along $C$, then the fiber types are either $(IV,III)$ or $(III,II)$.
\end{itemize}
To see this, let the seven-brane with fibers $F_1$
and $F_2$ be localized on the divisors $z=0$ and $t=0$ respectively. Then write 
\begin{equation}
f = z^a t^b \, F \qquad g = z^c t^d\, G
\end{equation}
where $a,b,c,d$ can be determined from Table \ref{tab:fibs} given the knowledge of $F_1$
and $F_2$. The discriminant takes the form
\begin{equation}
\Delta = z^{min(3a,2c)}\, t^{min(3b,2d)} \,\, \tilde \Delta
\end{equation}
and there is an extra brane along $C$ whenever $\tilde\Delta|_{z=t=0}=0$.
In the cases where there are no $(4,6)$ curves, the above conclusions can all be seen directly from Table
\ref{table:minimal models with extra brane}, where ``$NF$'' denotes that the $J$ invariant
of $I_0^*$ is not fixed.

\begin{table}
\centering
\begin{tabular}{ccccccc}
  $F_1$ & $F_2$ & $J_1$ & $J_2$ & Minimal on $C$& $\Delta$ & Additional Brane? \\ \hline \hline
$IV^*$ & $II$ & $0$ & $0$ & Yes & $(4 \tilde f^3\, t\, z + 27 \tilde g^2)\, t^2\, z^8$ & No \\
$I_0^*$ & $IV$ & $NF$ & $0$ & Yes & $(4 \tilde f^3\, t^2 + 27 \tilde g^2)\, t^4\, z^6$ & No \\
$I_0^*$ & $III$ & $NF$ & $1$ & Yes & $(4 \tilde f^3 +  27\tilde g^2 \, t ) \, t^3\, z^6$ & No \\
$I_0^*$ & $II$ & $NF$ & $0$ & Yes & $(4 \tilde f^3\, t + 27 \tilde g^2)\, t^2\, z^6$ & No \\
$IV$ & $IV$ & $0$ & $0$ & Yes & $(4 \tilde f^3\, t^2\, z^2 + 27 \tilde g^2)\, t^4\, z^4$ & No \\
$IV$ & $III$ & $0$ & $1$ & Yes & $(4 \tilde f^3\, z^2 + 27 \tilde g^2\, t)\, t^3\, z^4$ & Yes \\
$IV$ & $II$ & $0$ & $0$ & Yes & $(4 \tilde f^3\, t\, z^2 + 27 \tilde g^2)\, t^2\, z^4$ & No \\
$III$ & $III$ & $1$ & $1$ & Yes & $(4\tilde f^3 + 27\tilde g^2 \, t\, z)\, t^3\, z^3$ & No \\
$III$ & $II$ & $1$ & $0$ & Yes & $(4 \tilde f^3\, t + 27 \tilde g^2\, z)\, t^2\, z^3$ & Yes \\
$II$ & $II$ & $0$ & $0$ & Yes & $(4 \tilde f^3\, t\, z + 27 \tilde g^2)\, t^2\, z^2$ & No \\ \hline \hline
\end{tabular}
\caption{The possible intersecting elliptic seven-branes that may arise, according to their singular fibers $F_1$ and $F_2$, and some of their
properties.}
\label{table:minimal models with extra brane}
\end{table}

\subsection{Associated Matter Representations}

There are two possible fiber intersections that force the existence
of an extra brane, $IV$-$III$ and $III$-$II$. Due to the effects of outer
monodromy, the $IV$-$III$ collision allows for the possibility of
intersecting non-abelian seven-branes with $SU(3)\times SU(2)$ or
$SU(2) \times SU(2)$ gauge symmetry, while intersecting seven-branes
with a fiber collision $IV$-$III$ necessarily have $SU(2)$ gauge
symmetry. The Lie algebra representations of matter at the $IV$-$III$-$I_1$
$SU(3)\times SU(2)$ collision was determined in \rcite{GrassiHalversonShanesonTaylor2014}. They are
hypermultiplets of 
\begin{equation}
(3,2), (3,1), (1,2)
\end{equation} 
in the absence of flux, but can become the chiral non-abelian $SU(3)\times SU(2)$
representations of the standard model if chirality inducing G-flux is turned on.

Let us determine the Lie algebra representations in the
case of the other collision, which is $III$-$II$-$I_1$ via
anomaly cancellation in six dimensions. 
Consider a six-dimensional F-theory compactification with $B_2=\bP^2$
and a seven-brane with $SU(2)$ gauge symmetry from a type $III$ fiber on
$Z=\{z=0\}$, a divisor in the hyperplane class. Take also a seven-brane with
no gauge symmetry and a type $II$ singular fiber on $T=\{t=0\}$, also
in the hyperplane class. Such a Weierstrass model takes the form
\eqn{f_{12} = z \, t\, f_{10} \qquad \qquad g_{18} = z^2 \, t \,g_{15}
  \qquad \qquad \Delta = z^3 t^2\, (4f_{10}^3\, t+27g_{15}^2\,
  z)\equiv z^3 \, \tilde \Delta} and we would like to study matter at
the intersection $z = \tilde \Delta = 0$. These intersections are of
two types: the single point $z=t=0$ and the $10$ points $z=f_{10}=0$
with $t\ne 0$. The latter are all points where seven-branes with a
type $III$ and $I_1$ fiber collide; there are $2$ fundamentals of
$SU(2)$ for each such point \cite{Grassi:2000we,Grassi:2011hq}. Thus, the $10$ points contribute $20$
fundamental hypermultiplets. Since anomaly cancellation for an $SU(2)$
seven-brane on the hyperplane in $\bP^2$ requires $22$ hypermultiplets
(see e.g. section 2.5 of \cite{Johnson:2014xpa}), anomaly cancellation
requires that the $III$-$II$-$I_1$ intersection also contribute two
fundamental hypermultiplets.

\newsec{Argyres-Douglas Theories on D3-branes and BPS Dyons as Junctions}
\label{sec:D3 probes}

D3-branes in the vicinity of seven-branes realize non-trivial $\cN=2$
quantum field theories on their worldvolume \cite{Banks:1996nj}, which
are broken to $\cN=1$ theories by the background. The position of the
D3-brane relative to a seven-brane determines a point on the Coulomb
branch of the D3-brane theory, and the seven-brane determines a singular
point on the Coulomb branch at which additional particle states become
massless. This point is reached when the D3-brane collides with the
seven-brane, and the additional light states are string junctions that
stretch from the seven-brane to the D3-brane. 
Previous works \cite{Mikhailov:1998bx,DeWolfe:1998bi}
determined the string junction representations of BPS particles
for well-known $\cN=2$ theories, including $N_f=0,1,2,3,4$
Seiberg-Witten theory \cite{Seiberg:1994rs,Seiberg:1994aj}.

In this section we will do the same for D3-branes probing the type
$II$, $III$, and $IV$ singularities. Recall that the latter two are
associated with the non-perturbative $SU(2)$ and $SU(3)$ theories on
seven-branes that were discussed in section \ref{sec:su3su2}. Relative
to weakly coupled realizations of $SU(2)$ and $SU(3)$ from
$2$ $D7$-branes and 3 $D7$-branes, which have an $I_2$ and $I_3$ fiber in
F-theory, the type $III$ and $IV$ theories have an extra seven-brane. 
Via studying local axiodilaton profiles we have seen that in the vicinity of the brane
the string coupling is $O(1)$ and that these seven-branes source a nilpotent
$SL(2,\bZ)$ monodromy. Therefore, the worldvolume theories of nearby D3-branes necessarily
differ from the $N_f=2,3$ Seiberg-Witten theories realized on D3-branes near $I_2$ and $I_3$ seven-branes.

The worldvolume theory on a D3-brane when it has collided with a type $III$ or type $IV$ seven-brane is the
$N_f=2$ or $3$ Seiberg-Witten theory at its Argyres-Douglas point, respectively. 
We will see that the characteristic \cite{Argyres:1995jj} Argyres-Douglas (AD) phenomenon is
realized by string junctions. This phenomenon is
existence of points in the moduli space where electrons and dyons
charged under the $U(1)$ of the $\cN=2$ theory simultaneously become massless. In
F-theory this occurs when the D3-brane collides with a type $III$ or
type $IV$ seven-brane. The AD phenomenon was originally realized in
$G=SU(3)$ theories, which have genus two Seiberg-Witten curves, and thus one
might expect it to not exist in an elliptically fibered setup such as F-theory.
For specially tuned $G=SU(2)$ Seiberg-Witten theories Argyres-Douglas
points do exist \cite{Argyres:1995xn}, though; the tuning brings in an extra singularity
that turns a type $I_2$ ($I_3$) fiber into a type $III$ ($IV$) fiber. Somewhat ironically,
F-theory models with non-Higgsable seven-branes of type $III$ and $IV$ do not require any
such tuning; the topology of the base $B$ forces the local structure of the Seiberg-Witten
curve that would appear tuned from an $\cN=2$ point of view.

We will use the BPS junction criterion\footnote{See
  \cite{Mikhailov:1998bx} for another study of BPS junctions.} of
\cite{DeWolfe:1998bi} to determine the BPS states on three-branes near
the type $II$, $III$, and $IV$ seven-branes. This constraint is
\begin{equation}
\label{eq:BPS}
(J,J) \geq -2 + gcd(a(J))
\end{equation}
where $J$ is the string junction and $a(J)$ is its asymptotic charge,
and it is derived from the requirement that in the M-theory picture
$J$ is a holomorphic curve with boundary, with the boundary a
non-trivial one-cycle $a(J)$ in the smooth elliptic curve above the
D3-brane. This one-cycle $a(J)$ is the asymptotic charge, and it
determines the electric and magnetic charge of the associated junction
under the $U(1)$ of the D3-brane theory. Junctions are relative
homology cycles \cite{Grassi:2014ffa}, i.e. two-cycles that may have
boundary in a smooth elliptic curve above a particular point $p$. This
point is given physical meaning if $p$ is the location of a D3-brane.

Throughout this section we will need a basis choice in order to represent the one-cycle $a(J)$ as a vector in
$\bZ^2$. We choose
\begin{equation}
\pi_1 = \begin{pmatrix}1\\ 0 \end{pmatrix}, \qquad \pi_3 = \begin{pmatrix}0 \\ 1\end{pmatrix},
\end{equation}
which determines $\pi_2$ via $\pi_1+\pi_2+\pi_3=0\in H_1(E_p,\bZ)$, where $E_p$ is the elliptic fiber above
the D3-brane in the F-theory picture, or alternatively the torus that defines the electric and magnetic
charges.

\subsection{The $N_f=1$ Argyres-Douglas Point and the Type $II$ Kodaira Fiber}

Let us study the F-theory realization of BPS states on a D3-brane near a slightly deformed type $II$ Kodaira fiber. From \cite{Grassi:2014sda},
the vanishing cycles are $Z=\{\pi_3,\pi_1\}$ and so the $I$-matrix that determines topological intersections is
\begin{equation}
I = (\cdot,\cdot) = \begin{pmatrix} -1 & -\frac12 \\ -\frac12 & -1\end{pmatrix}.
\end{equation}
Defining $J=(Q_1,Q_2)$ we have
\begin{equation}
(J,J) = -Q_1Q_2-Q_1^2-Q_2^2 \qquad \text{with}\qquad a(J) = \begin{pmatrix} Q_2 \\ Q_1 \end{pmatrix}.
\end{equation}
The string junctions satisfying the BPS particle condition \eqref{eq:BPS} are
\begin{center}
\begin{tabular}{c|c}
$a(J)$ & Junctions \\ \hline
$(1,0)$ & $(0,1)$ \\
$(1,-1)$ & $(-1,1)$ \\
$(0,1)$ & $(1,0)$
\end{tabular}
\end{center}
These BPS particles arising from string junctions also have BPS
anti-particles via the action $J\mapsto -J$, which preserves $(J,J)$
and $gcd(a(J))$ and therefore the associated junctions $-J$ satisfy
\eqref{eq:BPS}. In the M-theory description the geometric object $J$
corresponds to a two-cycle, and the associated BPS particle arises
from wrapping an M2-brane on that cycle; the anti-particle associated
to $-J$ arises from wrapping an anti M2-brane.  Note that there is no
junction $J$ with $a(J)=0$ and $(J,J)=-2$; as explained in
\cite{Grassi:2014sda}, this demonstrates via deformation that the type
II singularity does not carry a gauge algebra.

One might be tempted to think that the seven-brane associated to a type II singularity has little impact on the 
low energy physics of an F-theory compactification, since it does not carry any gauge algebra. However, this
is not true, and we would like to emphasize:
\begin{itemize}
\item Locally near the type II seven-brane (or precisely in the 8d theory) the worldvolume theory on the D3-brane is
  $N_f=1$ Seiberg-Witten theory near its Argyres-Douglas point. At that point BPS electrons, monopoles, and dyons 
  become massless.
\item Since $J=0$ for the type $II$ Kodaira fiber, $g_s$ is $O(1)$ in the vicinity of the seven-brane,
  with $\tau$ profile solved via Ramanujan's alternative bases in section \ref{sec:axiodilaton}..
\item Even though both can split into two mutually non-local seven-branes, the seven-brane associated to a 
type II Kodaira fiber is not an orientifold. First, because their $SL(2,\bZ)$ monodromies are different, and second because the orientifold famously \emph{must} split due
to instanton effects in F-theory, which is not true of the type II seven-brane.
\end{itemize}

\subsection{The $N_f=2$ Argyres-Douglas Point and the Type $III$ Kodaira Fiber}

In this section we study the F-theory realization of BPS states on D3-branes near a slightly deformed type $III$ Kodaira fiber.
In the coincident limit this theory is the Argyres-Douglas theory obtained by tuning $N_f=2$ $G=SU(2)$
Seiberg-Witten theory to its Argyres-Douglas point. 
Utilizing an explicit deformation of \cite{Grassi:2014sda}, the vanishing cycles are $Z=\{\pi_2,\pi_1,\pi_3\}$ and the 
$I$-matrix that determines topological intersections is 
\begin{equation}
I = (\cdot,\cdot) = \begin{pmatrix} -1 & \frac12 & -\frac12 \\ \frac12 & -1 & \frac12 \\ -\frac12 & \frac12 & -1  \end{pmatrix}.
\end{equation}
Defining a string junction by $J=(Q_1,Q_2,Q_3)$ we have 
\begin{equation}
(J,J) = Q_1Q_2+Q_1Q_3-Q_2Q_3 - \sum_i Q_i^2 \qquad \text{with}\qquad a(J) = \begin{pmatrix}-Q_1+Q_2 \\ -Q_1+Q_3 \end{pmatrix}.
\end{equation}
Thees
BPS particles arising from string junctions also have BPS anti-particles via the action $J\mapsto -J$, which
preserves $(J,J)$ and $gcd(a(J))$, leaving (\ref{eq:BPS}) invariant.

Using the constraint \eqref{eq:BPS} the possible BPS string junctions
can be computed directly. The states of self-intersection $-2$ have
$a(J)=0$ and were identified in \cite{Grassi:2014sda}. They are $J_+ = (1,1,1)$ and $J_-=(-1,-1,-1)$,
where the $\pm$ denote a choice of positive and negative root for the associated
$SU(2)$ algebra that is the gauge algebra on the seven-brane and the flavor
algebra of the three-brane theory. The rest of the junctions satisfying \eqref{eq:BPS} have
$(J,J)=-1$ and are given by
\begin{center}
\begin{tabular}{c|c}
$a(J)$ & Junctions \\ \hline
$(0,1)$ & $(0,0,1)$,$(-1,-1,0)$ \\
$(1,0)$ & $(0,1,0)$,$(-1,0,-1)$ \\
$(1,1)$ & $(-1,0,0)$ \\
$(1,-1)$ & $(0,1,-1)$
\end{tabular}
\end{center}
where the sets of junctions in the first two lines are doublets of
$SU(2)$ since they differ by $J_\pm$.  This spectrum matches the known
BPS states in the maximal chamber of the deformed theory with masses
turned on; see e.g. \cite{Maruyoshi:2013fwa}.

We have seen the relationship between the ordered sets of
vanishing cycles $Z=\{\pi_2,\pi_1,\pi_3\}$ of \cite{Grassi:2014sda}
and\footnote{The set of vanishing cycles $\{A,A,C\}$ should be compared to the set $\{\pi_1,\pi_1,\pi_3\}$ of section \ref{sec:su3su2}.} $Z=\{A,A,C\}$ of \cite{DeWolfeZwiebach,DeWolfe:1998bi} that can be associated to the
deformed type III Kodaira fiber by explicit
deformation in section \ref{sec:su3su2}. Let us study the BPS states with the
latter set for the sake of completeness, taking
\begin{equation}
A=\begin{pmatrix}1 \\ 0\end{pmatrix}, \qquad C = \begin{pmatrix}1 \\ 1\end{pmatrix},
\end{equation}
as in \cite{DeWolfe:1998zf}.
 The $I$-matrix is
\begin{equation}
I = (\cdot,\cdot) = \begin{pmatrix} -1 & 0 & \frac12 \\ 0 & -1 & \frac12 \\ \frac12 & \frac12 & -1  \end{pmatrix}
\end{equation}
and taking $J=(Q_1,Q_2,Q_3)$ we compute
\begin{equation}
(J,J) = Q_3(Q_1+Q_2) - \sum_i Q_i^2 \qquad \text{with}\qquad a(J) = \begin{pmatrix}Q_1+Q_2+Q_3\\ Q_3 \end{pmatrix}.
\end{equation}
There are junctions $J_\pm$ satisfying the BPS constraint \eqref{eq:BPS} with $a(J)=0$ and $(J,J)=-2$. They are
$J_+=(1,-1,0)$ and $J_-=(-1,1,0)$, and they are the $W_\pm$ bosons of the broken $SU(2)$ theory. The rest of the
junctions satisfying the BPS condition have $(J,J)=-1$ and satisfy
\begin{center}
\begin{tabular}{c|c}
$a(J)$ & Junctions \\ \hline
$(3,1)$ & $(1,1,1)$ \\
$(2,1)$ & $(1,0,1)$,$(0,1,1)$ \\
$(1,1)$ & $(1,0,1)$ \\
$(1,0)$ & $(1,0,0)$,$(0,1,0)$.
\end{tabular}
\end{center}
We again see two $SU(2)$ doublets and two $SU(2)$ singlets before taking into account the $-J$ junctions.
These junctions with electromagnetic charge 
must be related to those of $Z=\{\pi_2,\pi_1,\pi_3\}$ by a Hanany-Witten move, as we explicitly showed for the simple roots
$J_\pm$ in section \ref{sec:su3su2}.

\subsection{The $N_f=3$ Argyres-Douglas Point and the Type $IV$ Kodaira Fiber}

In this section we study the F-theory realization of BPS states on D3-branes near a slightly deformed type
IV Kodaira fiber. In the coincident limit the worldvolume theory on the D3-brane is the
Argyres-Douglas theory obtained by tuning $N_f=3$ $G=SU(2)$ Seiberg-Witten theory to
its Argyres-Douglas point.

The ordered set of vanishing cycles derived in \cite{Grassi:2014sda} is $Z_{IV}=\{\pi_1,\pi_3,\pi_1,\pi_3\}$
and
the associated I-matrix that determines the topological intersections of junctions is
\begin{equation}
I = (\cdot,\cdot) = \begin{pmatrix} -1 & \frac12 & 0 & \frac12 \\ \frac12 & -1 & -\frac12 & 0 \\ 0 & -\frac12 & -1 & \frac12 
\\ \frac12 & 0 & \frac12 & -1 \end{pmatrix}.
\end{equation}
Writing the junction as a vector $J=(Q_1,Q_2,Q_3,Q_4)$ we have 
\begin{equation}
(J,J) = Q_1Q_2 + Q_1Q_4-Q_2Q_3+Q_3Q_4 - \sum_i Q_i^2 \qquad \text{with}\qquad a(J) = \begin{pmatrix}Q_1+Q_3 \\ Q_2+Q_4 \end{pmatrix}.
\end{equation}
These
BPS particles arising from string junctions also have BPS anti-particles via the action $J\mapsto -J$, which
preserves $(J,J)$ and $a(J)$ and therefore the associated junctions $-J$ satisfy \eqref{eq:BPS}. The latter arise from
wrapped anti M2-branes in the M-theory picture.

Let us derive the possible BPS string junctions using the constraint \eqref{eq:BPS}.
The junctions of self-intersection $-2$ have
$a(J)=0$ and were identified in \cite{Grassi:2014sda}. They are the roots of an $SU(3)$ algebra,
and we take $J_1=(1,0,-1,0)$
and $J_2=(0,1,0,-1)$ as the simple roots; then the full set of root junctions is simply $J_1$,$J_2$,
$J_1+J_2$ and their negatives. Solving the condition for BPS junctions \eqref{eq:BPS} we find that the
possible BPS junctions are
\begin{center}
\begin{tabular}{c|c}
$a(J)$ & Junctions \\ \hline
$(-1,-1)$ & $(0,0,-1,-1)$  $(-1,0,0,-1)$  $(-1,-1,0,0)$ \\
$(1,0)$ & $(1,0,0,0)$  $(0,0,1,0)$  $(0,-1,1,1)$ \\
$(0,1)$ & $(1,1,-1,0)$ $(0,1,0,0)$ $(0,0,0,1)$ \\
$(2,1)$ & $(1,0,1,1)$ \\
$(-1,-2)$ & $(-1,-1,0,-1)$ \\
$(-1,1)$ & $(0,1,-1,0)$   
\end{tabular}
\end{center}
and we have chosen the ordering of the junctions in the first three
columns to show that they are fundamentals of $SU(3)$. Namely, in each
of the first thee columns the second junction subtracted from the
first is $J_1$ and the third junction from the second is $J_2$. The
negatives of these are anti-fundamental, completing the non-trivial
flavor hypermultiplets.

\acknowledgments \vspace{.25cm} I would like to thank Philip Argyres,
Shaun Cooper, Andreas Malmendier, Brent Nelson, Daniel Schultz,
Washington Taylor, Yi-Nan Wang and Wenbin Yan for helpful discussions
and correspondence. I am particularly grateful to Antonella Grassi and
Julius Shaneson for discussions and comments on a draft, and to
J.L. Halverson for support and encouragement.  This work is generously
supported by startup funding from Northeastern University and the
National Science Foundation under Grant No. PHY11-25915.

\begin{table}
\centering
\begin{tabular}{ccccccc}
  $F_1$ & $F_2$ & $J_1$ & $J_2$ & Minimal on $C$& $\Delta$ & Additional Brane? \\ \hline \hline 
$II^*$ & $II^*$ & $0$ & $0$ & No & $(4 \tilde f^3\, t^2\, z^2 + 27 \tilde g^2)\, t^{10}\, z^{10}$ & No \\
$II^*$ & $III^*$ & $0$ & $1$ & No & $(4 \tilde f^3\, z^2 + 27 \tilde g^2\, t)\, t^9\, z^{10}$ & Yes \\
$II^*$ & $IV^*$ & $0$ & $0$ & No & $(4 \tilde f^3\, t\, z^2 + 27 \tilde g^2)\, t^8\, z^{10}$ & No \\
$II^*$ & $I_0^*$ & $0$ & $NF$ & No & $(4 \tilde f^3\, z^2 + 27 \tilde g^2)\, t^6\, z^{10}$ & No\\
$II^*$ & $IV$ & $0$ & $0$ & No & $(4 \tilde f^3\, t^2\, z^2 + 27 \tilde g^2)\, t^4\, z^{10}$ & No \\
$II^*$ & $III$ & $0$ & $1$ & No & $(4 \tilde f^3\, z^2 + 27 \tilde g^2\, t)\, t^3\, z^{10}$ & Yes \\
$II^*$ & $II$ & $0$ & $0$ & No & $(4 \tilde f^3\, t\, z^2 + 27 \tilde g^2)\, t^2\, z^{10}$ & No \\
$III^*$ & $III^*$ & $1$ & $1$ & No & $(4 \tilde f^3 + 27\tilde g^2 \, t)\, t^9\, z^9$ & No \\
$III^*$ & $IV^*$ & $1$ & $0$ & No & $(4 \tilde f^3\, t + 27 \tilde g^2\, z)\, t^8\, z^9$ & Yes \\
$III^*$ & $I_0^*$ & $1$ & $NF$ & No & $(4 \tilde f^3 + 27\tilde g^2 \, z)\, t^6 \,z^9$ & No \\
$III^*$ & $IV$ & $1$ & $0$ & No & $(4 \tilde f^3\, t^2 + 27 \tilde g^2\, z)\, t^4\, z^9$ & Yes \\
$III^*$ & $III$ & $1$ & $1$ & No & $(4\tilde f^3 + 27\tilde g^2 \, t\, z)\, t^3\, z^9$ & No \\
$III^*$ & $II$ & $1$ & $0$ & No & $(4 \tilde f^3\, t + 27 \tilde g^2\, z)\, t^2\, z^9$ & Yes \\
$IV^*$ & $IV^*$ & $0$ & $0$ & No & $(4 \tilde f^3\, t\, z + 27 \tilde g^2)\, t^8\, z^8$ & No \\
$IV^*$ & $I_0^*$ & $0$ & $NF$ & No & $(4 \tilde f^3\, z + 27 \tilde g^2)\, t^8 z^8$ & No \\
$IV^*$ & $IV$ & $0$ & $0$ & No & $(4 \tilde f^3\, t^2\, z + 27 \tilde g^2)\, t^4\, z^8$ & No \\
$IV^*$ & $III$ & $0$ & $1$ & No & $(4\tilde f^3\, z + 27\tilde g^2 \, t) \, t^3\, z^8$ & Yes \\
$IV^*$ & $II$ & $0$ & $0$ & Yes & $(4 \tilde f^3\, t\, z + 27 \tilde g^2)\, t^2\, z^8$ & No \\
$I_0^*$ & $I_0^*$ & $NF$ & $NF$ & No & $(4 \tilde f^3 + 27 \tilde g^2) \, t^6\, z^6$ & No \\
$I_0^*$ & $IV$ & $NF$ & $0$ & Yes & $(4 \tilde f^3\, t^2 + 27 \tilde g^2)\, t^4\, z^6$ & No \\
$I_0^*$ & $III$ & $NF$ & $1$ & Yes & $(4 \tilde f^3 +  27\tilde g^2 \, t ) \, t^3\, z^6$ & No \\
$I_0^*$ & $II$ & $NF$ & $0$ & Yes & $(4 \tilde f^3\, t + 27 \tilde g^2)\, t^2\, z^6$ & No \\
$IV$ & $IV$ & $0$ & $0$ & Yes & $(4 \tilde f^3\, t^2\, z^2 + 27 \tilde g^2)\, t^4\, z^4$ & No \\
$IV$ & $III$ & $0$ & $1$ & Yes & $(4 \tilde f^3\, z^2 + 27 \tilde g^2\, t)\, t^3\, z^4$ & Yes \\
$IV$ & $II$ & $0$ & $0$ & Yes & $(4 \tilde f^3\, t\, z^2 + 27 \tilde g^2)\, t^2\, z^4$ & No \\
$III$ & $III$ & $1$ & $1$ & Yes & $(4\tilde f^3 + 27\tilde g^2 \, t\, z)\, t^3\, z^3$ & No \\
$III$ & $II$ & $1$ & $0$ & Yes & $(4 \tilde f^3\, t + 27 \tilde g^2\, z)\, t^2\, z^3$ & Yes \\
$II$ & $II$ & $0$ & $0$ & Yes & $(4 \tilde f^3\, t\, z + 27 \tilde g^2)\, t^2\, z^2$ & No \\ \hline\hline
\end{tabular}
\caption{Properties of all intersections of Kodaira fibers with finite order monodromy, including
those that give rise to $(4,6)$ curves. There is an additional brane if and only if $J_1+J_2=1$, and therefore
it may play an important role in local axiodilaton profiles.}
\end{table}

\bibliography{chetdocbib}

\begin{thebibliography}{10}
\ifx\href\asklfhas\newcommand{\href}[2]{#2}\fi
\ifx\arxivref\asklfhas\newcommand{\arxivref}[2]{\href{http://arxiv.org/abs/#1}{#2}}\fi
\ifx\doiref\asklfhas\newcommand{\doiref}[2]{\href{http://dx.doi.org/#1}{#2}}\fi
\parskip 0pt
\normalsize

\bibitem{V1}
C.~Vafa,
\textit{``Evidence for {$F$}-theory''},
\doiref{10.1016/0550-3213(96)00172-1}{Nuclear~Phys.~B \textbf{469}, 403
  (1996)},
\href{http://dx.doi.org/10.1016/0550-3213(96)00172-1}{\texttt{http://dx.doi.org/10.1016/0550-3213(96)00172-1}}.

\bibitem{GaberdielZwiebach}
M.~R. Gaberdiel \& B.~Zwiebach,
\textit{``Exceptional groups from open strings''},
\doiref{10.1016/S0550-3213(97)00841-9}{Nuclear~Phys.~B \textbf{518}, 151
  (1998)},
\href{http://dx.doi.org/10.1016/S0550-3213(97)00841-9}{\texttt{http://dx.doi.org/10.1016/S0550-3213(97)00841-9}}.

\bibitem{DeWolfe:1998zf}
O.~DeWolfe \& B.~Zwiebach,
\textit{``{String junctions for arbitrary Lie algebra representations}''},
\doiref{10.1016/S0550-3213(98)00743-3}{Nucl.~Phys. \textbf{B541}, 509 (1999)},
\normalsize{\texttt{\arxivref{hep-th/9804210}{hep-th/9804210}}}.

\bibitem{Grassi:2013kha}
A.~Grassi, J.~Halverson \& J.~L. Shaneson,
\textit{``{Matter From Geometry Without Resolution}''},
\doiref{10.1007/JHEP10(2013)205}{JHEP \textbf{1310}, 205 (2013)},
\normalsize{\texttt{\arxivref{1306.1832}{arXiv:1306.1832}}}.

\bibitem{Grassi:2014sda}
A.~Grassi, J.~Halverson \& J.~L. Shaneson,
\textit{``{Non-Abelian Gauge Symmetry and the Higgs Mechanism in F-theory}''},
\doiref{10.1007/s00220-015-2313-0}{Commun.~Math.~Phys. \textbf{336}, 1231
  (2015)},
\normalsize{\texttt{\arxivref{1402.5962}{arXiv:1402.5962}}}.

\bibitem{Grassi:2014ffa}
A.~Grassi, J.~Halverson \& J.~L. Shaneson,
\textit{``{Geometry and Topology of String Junctions}''},
\normalsize{\texttt{\arxivref{1410.6817}{arXiv:1410.6817}}}.

\bibitem{Braun:2014xka}
A.~P. Braun \& T.~Watari,
\textit{``{The Vertical, the Horizontal and the Rest: anatomy of the middle
  cohomology of Calabi-Yau fourfolds and F-theory applications}''},
\doiref{10.1007/JHEP01(2015)047}{JHEP \textbf{1501}, 047 (2015)},
\normalsize{\texttt{\arxivref{1408.6167}{arXiv:1408.6167}}}.

\bibitem{Watari:2015ysa}
T.~Watari,
\textit{``{Statistics of Flux Vacua for Particle Physics}''},
\normalsize{\texttt{\arxivref{1506.08433}{arXiv:1506.08433}}}.

\bibitem{MV}
D.~R. Morrison \& C.~Vafa,
\textit{``Compactifications of {$F$}-theory on {C}alabi-{Y}au threefolds. {I},
  II''},
Nuclear~Phys.~B \textbf{473, 476}, 74 (1996).

\bibitem{GrassiHalversonShanesonTaylor2014}
A.~Grassi, J.~Halverson, J.~Shaneson \& W.~Taylor,
\textit{``Non-{H}iggsable {QCD} and the Standard Model Spectrum in
  {F}-theory''},
{\tt arXiv::1409.8295}.

\bibitem{Cecotti:2010bp}
S.~Cecotti, C.~Cordova, J.~J. Heckman \& C.~Vafa,
\textit{``{T-Branes and Monodromy}''},
\doiref{10.1007/JHEP07(2011)030}{JHEP \textbf{1107}, 030 (2011)},
\normalsize{\texttt{\arxivref{1010.5780}{arXiv:1010.5780}}}.

\bibitem{Morrison:2012js}
D.~R. Morrison \& W.~Taylor,
\textit{``{Toric bases for 6D F-theory models}''},
\doiref{10.1002/prop.201200086}{Fortsch.~Phys. \textbf{60}, 1187 (2012)},
\normalsize{\texttt{\arxivref{1204.0283}{arXiv:1204.0283}}}.

\bibitem{Morrison:2012np}
D.~R. Morrison \& W.~Taylor,
\textit{``{Classifying bases for 6D F-theory models}''},
\doiref{10.2478/s11534-012-0065-4}{Central~Eur.~J.~Phys. \textbf{10}, 1072
  (2012)},
\normalsize{\texttt{\arxivref{1201.1943}{arXiv:1201.1943}}}.

\bibitem{Anderson:2014gla}
L.~B. Anderson \& W.~Taylor,
\textit{``{Geometric constraints in dual F-theory and heterotic string
  compactifications}''},
\doiref{10.1007/JHEP08(2014)025}{JHEP \textbf{1408}, 025 (2014)},
\normalsize{\texttt{\arxivref{1405.2074}{arXiv:1405.2074}}}.

\bibitem{Morrison:2014lca}
D.~R. Morrison \& W.~Taylor,
\textit{``{Non-Higgsable clusters for 4D F-theory models}''},
\doiref{10.1007/JHEP05(2015)080}{JHEP \textbf{1505}, 080 (2015)},
\normalsize{\texttt{\arxivref{1412.6112}{arXiv:1412.6112}}}.

\bibitem{Halverson:2015jua}
J.~Halverson \& W.~Taylor,
\textit{``{$ {\mathrm{\mathbb{P}}}^1 $-bundle bases and the prevalence of
  non-Higgsable structure in 4D F-theory models}''},
\doiref{10.1007/JHEP09(2015)086}{JHEP \textbf{1509}, 086 (2015)},
\normalsize{\texttt{\arxivref{1506.03204}{arXiv:1506.03204}}}.

\bibitem{Taylor:2015ppa}
W.~Taylor \& Y.-N. Wang,
\textit{``{A Monte Carlo exploration of threefold base geometries for 4d
  F-theory vacua}''},
\doiref{10.1007/JHEP01(2016)137}{JHEP \textbf{1601}, 137 (2016)},
\normalsize{\texttt{\arxivref{1510.04978}{arXiv:1510.04978}}}.

\bibitem{Na88}
N.~Nakayama,
\textit{``On {W}eierstrass models''},
in \textit{``Algebraic geometry and commutative algebra, Vol.\ II''},
Kinokuniya (1988),
Tokyo,
p.~405--431.

\bibitem{MR0187255}
K.~Kodaira,
\textit{``On the structure of compact complex analytic surfaces. {I}''},
Amer.~J.~Math. \textbf{86}, 751 (1964).

\bibitem{MR0205280}
K.~Kodaira,
\textit{``On the structure of compact complex analytic surfaces. {II}''},
Amer.~J.~Math. \textbf{88}, 682 (1966).

\bibitem{MR0228019}
K.~Kodaira,
\textit{``On the structure of compact complex analytic surfaces. {III}''},
Amer.~J.~Math. \textbf{90}, 55 (1968).

\bibitem{Morrison:1996pp}
D.~R. Morrison \& C.~Vafa,
\textit{``{Compactifications of F theory on Calabi-Yau threefolds. 2.}''},
\doiref{10.1016/0550-3213(96)00369-0}{Nucl.~Phys. \textbf{B476}, 437 (1996)},
\normalsize{\texttt{\arxivref{hep-th/9603161}{hep-th/9603161}}}.

\bibitem{Taylor:2015xtz}
W.~Taylor \& Y.-N. Wang,
\textit{``{The F-theory geometry with most flux vacua}''},
\doiref{10.1007/JHEP12(2015)164}{JHEP \textbf{1512}, 164 (2015)},
\normalsize{\texttt{\arxivref{1511.03209}{arXiv:1511.03209}}}.

\bibitem{Ashok:2003gk}
S.~Ashok \& M.~R. Douglas,
\textit{``{Counting flux vacua}''},
\doiref{10.1088/1126-6708/2004/01/060}{JHEP \textbf{0401}, 060 (2004)},
\normalsize{\texttt{\arxivref{hep-th/0307049}{hep-th/0307049}}}.

\bibitem{Denef2004}
F.~Denef \& M.~R. Douglas,
\textit{``{Distributions of flux vacua}''},
\doiref{10.1088/1126-6708/2004/05/072}{JHEP \textbf{0405}, 072 (2004)},
\normalsize{\texttt{\arxivref{hep-th/0404116}{hep-th/0404116}}}.

\bibitem{Taylor2015}
W.~Taylor \& Y.-N. Wang,
\textit{``{A Monte Carlo exploration of threefold base geometries for 4d
  F-theory vacua}''},
\normalsize{\texttt{\arxivref{1510.04978}{arXiv:1510.04978}}}.

\bibitem{MR0099904}
S.~Ramanujan,
\textit{``Notebooks. {V}ols. 1, 2''},
Tata Institute of Fundamental Research, Bombay (1957),
p.~Vol. 1. vi+351 pp.; Vol. 2. vi+393.

\bibitem{MR1311903}
B.~C. Berndt, S.~Bhargava \& F.~G. Garvan,
\textit{``Ramanujan's theories of elliptic functions to alternative bases''},
\doiref{10.2307/2155035}{Trans.~Amer.~Math.~Soc. \textbf{347}, 4163 (1995)},
\href{http://dx.doi.org/10.2307/2155035}{\texttt{http://dx.doi.org/10.2307/2155035}}.

\bibitem{MR1117903}
B.~C. Berndt,
\textit{``Ramanujan's notebooks. {P}art {III}''},
Springer-Verlag, New York (1991),
p.~xiv+510.

\bibitem{MR1071759}
J.~M. Borwein \& P.~B. Borwein,
\textit{``A remarkable cubic mean iteration''},
in \textit{``Computational methods and function theory ({V}alpara\'\i so,
  1989)''},
Springer, Berlin (1990),
p.~27--31,
\href{http://dx.doi.org/10.1007/BFb0087894}{\texttt{http://dx.doi.org/10.1007/BFb0087894}}.

\bibitem{MR1010408}
J.~M. Borwein \& P.~B. Borwein,
\textit{``A cubic counterpart of {J}acobi's identity and the {AGM}''},
\doiref{10.2307/2001551}{Trans.~Amer.~Math.~Soc. \textbf{323}, 691 (1991)},
\href{http://dx.doi.org/10.2307/2001551}{\texttt{http://dx.doi.org/10.2307/2001551}}.

\bibitem{MR1237931}
J.~Borwein, P.~Borwein \& F.~Garvan,
\textit{``Hypergeometric analogues of the arithmetic-geometric mean
  iteration''},
\doiref{10.1007/BF01204654}{Constr.~Approx. \textbf{9}, 509 (1993)},
\href{http://dx.doi.org/10.1007/BF01204654}{\texttt{http://dx.doi.org/10.1007/BF01204654}}.

\bibitem{MR1243610}
J.~M. Borwein, P.~B. Borwein \& F.~G. Garvan,
\textit{``Some cubic modular identities of {R}amanujan''},
\doiref{10.2307/2154520}{Trans.~Amer.~Math.~Soc. \textbf{343}, 35 (1994)},
\href{http://dx.doi.org/10.2307/2154520}{\texttt{http://dx.doi.org/10.2307/2154520}}.

\bibitem{MR1825995}
B.~C. Berndt, H.~H. Chan \& W.-C. Liaw,
\textit{``On {R}amanujan's quartic theory of elliptic functions''},
\doiref{10.1006/jnth.2000.2615}{J.~Number~Theory \textbf{88}, 129 (2001)},
\href{http://dx.doi.org/10.1006/jnth.2000.2615}{\texttt{http://dx.doi.org/10.1006/jnth.2000.2615}}.

\bibitem{MR3107523}
D.~Schultz,
\textit{``Cubic theta functions''},
\doiref{10.1016/j.aim.2013.08.021}{Adv.~Math. \textbf{248}, 618 (2013)},
\href{http://dx.doi.org/10.1016/j.aim.2013.08.021}{\texttt{http://dx.doi.org/10.1016/j.aim.2013.08.021}}.

\bibitem{Cooper2009}
S.~Cooper,
\textit{``Inversion formulas for elliptic functions''},
\doiref{10.1112/plms/pdp007}{Proc.~Lond.~Math.~Soc.~(3) \textbf{99}, 461
  (2009)},
\href{http://dx.doi.org/10.1112/plms/pdp007}{\texttt{http://dx.doi.org/10.1112/plms/pdp007}}.

\bibitem{Sen:1996vd}
A.~Sen,
\textit{``{F theory and orientifolds}''},
\doiref{10.1016/0550-3213(96)00347-1}{Nucl.~Phys. \textbf{B475}, 562 (1996)},
\normalsize{\texttt{\arxivref{hep-th/9605150}{hep-th/9605150}}}.

\bibitem{Sen:1997gv}
A.~Sen,
\textit{``{Orientifold limit of F theory vacua}''},
\doiref{10.1103/PhysRevD.55.R7345}{Phys.~Rev. \textbf{D55}, 7345 (1997)},
\normalsize{\texttt{\arxivref{hep-th/9702165}{hep-th/9702165}}}.

\bibitem{Dasgupta:1996ij}
K.~Dasgupta \& S.~Mukhi,
\textit{``{F theory at constant coupling}''},
\doiref{10.1016/0370-2693(96)00875-1}{Phys.~Lett. \textbf{B385}, 125 (1996)},
\normalsize{\texttt{\arxivref{hep-th/9606044}{hep-th/9606044}}}.

\bibitem{Cvetic:2010rq}
M.~Cvetic, I.~Garcia-Etxebarria \& J.~Halverson,
\textit{``{Global F-theory Models: Instantons and Gauge Dynamics}''},
\doiref{10.1007/JHEP01(2011)073}{JHEP \textbf{1101}, 073 (2011)},
\normalsize{\texttt{\arxivref{1003.5337}{arXiv:1003.5337}}}.

\bibitem{DeWolfeZwiebach}
O.~DeWolfe \& B.~Zwiebach,
\textit{``String junctions for arbitrary {L}ie-algebra representations''},
\doiref{10.1016/S0550-3213(98)00743-3}{Nuclear~Phys.~B \textbf{541}, 509
  (1999)},
\href{http://dx.doi.org/10.1016/S0550-3213(98)00743-3}{\texttt{http://dx.doi.org/10.1016/S0550-3213(98)00743-3}}.

\bibitem{Grassi:2000we}
A.~Grassi \& D.~R. Morrison,
\textit{``{Group representations and the Euler characteristic of elliptically
  fibered Calabi-Yau threefolds}''},
\normalsize{\texttt{\arxivref{math/0005196}{math/0005196}}}.

\bibitem{Grassi:2011hq}
A.~Grassi \& D.~R. Morrison,
\textit{``{Anomalies and the Euler characteristic of elliptic Calabi-Yau
  threefolds}''},
\doiref{10.4310/CNTP.2012.v6.n1.a2}{Commun.~Num.~Theor.~Phys. \textbf{6}, 51
  (2012)},
\normalsize{\texttt{\arxivref{1109.0042}{arXiv:1109.0042}}}.

\bibitem{Johnson:2014xpa}
S.~B. Johnson \& W.~Taylor,
\textit{``{Calabi-Yau threefolds with large $h^{2,1}$}''},
\doiref{10.1007/JHEP10(2014)023}{JHEP \textbf{1410}, 23 (2014)},
\normalsize{\texttt{\arxivref{1406.0514}{arXiv:1406.0514}}}.

\bibitem{Banks:1996nj}
T.~Banks, M.~R. Douglas \& N.~Seiberg,
\textit{``{Probing F theory with branes}''},
\doiref{10.1016/0370-2693(96)00808-8}{Phys.~Lett. \textbf{B387}, 278 (1996)},
\normalsize{\texttt{\arxivref{hep-th/9605199}{hep-th/9605199}}}.

\bibitem{Mikhailov:1998bx}
A.~Mikhailov, N.~Nekrasov \& S.~Sethi,
\textit{``{Geometric realizations of BPS states in N=2 theories}''},
\doiref{10.1016/S0550-3213(98)80001-1}{Nucl.~Phys. \textbf{B531}, 345 (1998)},
\normalsize{\texttt{\arxivref{hep-th/9803142}{hep-th/9803142}}}.

\bibitem{DeWolfe:1998bi}
O.~DeWolfe, T.~Hauer, A.~Iqbal \& B.~Zwiebach,
\textit{``{Constraints on the BPS spectrum of N=2, D = 4 theories with A-D-E
  flavor symmetry}''},
\doiref{10.1016/S0550-3213(98)00652-X}{Nucl.~Phys. \textbf{B534}, 261 (1998)},
\normalsize{\texttt{\arxivref{hep-th/9805220}{hep-th/9805220}}}.

\bibitem{Seiberg:1994rs}
N.~Seiberg \& E.~Witten,
\textit{``{Electric - magnetic duality, monopole condensation, and confinement
  in N=2 supersymmetric Yang-Mills theory}''},
\doiref{10.1016/0550-3213(94)90124-4}{Nucl.~Phys. \textbf{B426}, 19 (1994)},
\normalsize{\texttt{\arxivref{hep-th/9407087}{hep-th/9407087}}},
[Erratum: Nucl. Phys.B430,485(1994)].

\bibitem{Seiberg:1994aj}
N.~Seiberg \& E.~Witten,
\textit{``{Monopoles, duality and chiral symmetry breaking in N=2
  supersymmetric QCD}''},
\doiref{10.1016/0550-3213(94)90214-3}{Nucl.~Phys. \textbf{B431}, 484 (1994)},
\normalsize{\texttt{\arxivref{hep-th/9408099}{hep-th/9408099}}}.

\bibitem{Argyres:1995jj}
P.~C. Argyres \& M.~R. Douglas,
\textit{``{New phenomena in SU(3) supersymmetric gauge theory}''},
\doiref{10.1016/0550-3213(95)00281-V}{Nucl.~Phys. \textbf{B448}, 93 (1995)},
\normalsize{\texttt{\arxivref{hep-th/9505062}{hep-th/9505062}}}.

\bibitem{Argyres:1995xn}
P.~C. Argyres, M.~R. Plesser, N.~Seiberg \& E.~Witten,
\textit{``{New N=2 superconformal field theories in four-dimensions}''},
\doiref{10.1016/0550-3213(95)00671-0}{Nucl.~Phys. \textbf{B461}, 71 (1996)},
\normalsize{\texttt{\arxivref{hep-th/9511154}{hep-th/9511154}}}.

\bibitem{Maruyoshi:2013fwa}
K.~Maruyoshi, C.~Y. Park \& W.~Yan,
\textit{``{BPS spectrum of Argyres-Douglas theory via spectral network}''},
\doiref{10.1007/JHEP12(2013)092}{JHEP \textbf{1312}, 092 (2013)},
\normalsize{\texttt{\arxivref{1309.3050}{arXiv:1309.3050}}}.

\end{thebibliography}
\end{document}